\documentclass[a4paper, 11pt]{article}

\usepackage{jcappub} 
\usepackage[utf8]{inputenc}
\usepackage[T1]{fontenc}

\usepackage[mathscr]{eucal}
\usepackage[Symbolsmallscale]{upgreek}
\usepackage{xcolor}
\usepackage{slashed}
\usepackage{hyperref}
\usepackage{placeins}
\usepackage{listings}
\usepackage{float}
\usepackage{listings}
\usepackage{cleveref}
\usepackage{fontawesome}
\usepackage{xcolor} 

\definecolor{codegreen}{rgb}{0, 0.6, 0}
\definecolor{codegray}{rgb}{0.5, 0.5, 0.5}
\definecolor{codepurple}{rgb}{0.58, 0, 0.82}
\definecolor{backcolour}{rgb}{0.95, 0.95, 0.92}
\definecolor{linkblue}{rgb}{0.0, 0.0, 0.5}

\hypersetup{
  colorlinks=true, 
  linkcolor=linkblue,   
  urlcolor=linkblue, 
  citecolor=linkblue
  }

\lstdefinestyle{mystyle}{
  backgroundcolor=\color{backcolour},  
  commentstyle=\color{codegreen}, 
  keywordstyle=\color{magenta}, 
  numberstyle=\tiny\color{codegray}, 
  stringstyle=\color{codepurple}, 
  basicstyle=\ttfamily\footnotesize, 
  breakatwhitespace=false,     
  breaklines=true,         
  captionpos=b,           
  keepspaces=true,         
  numbers=left,           
  numbersep=5pt,          
  showspaces=false,         
  showstringspaces=false, 
  showtabs=false,          
  tabsize=2
}

\definecolor{darkgreen}{rgb}{0.0, 0.5, 0.0}

\title{Worth the Effort? An Examination on the Effect of Higher Diligence Calculations of the Sound Shell Model}

\author[a]{Fazlollah Hajkarim, }
\author[b]{Graham White, }
\author[c]{Yang Xiao }

\affiliation[a]{Department of Physics and Astronomy, University of Oklahoma, Norman, OK 73019, USA}
\affiliation[b]{School of Physics and Astronomy, University of Southampton, Southampton SO17 1BJ, United Kingdom}
\affiliation[c]{School of Physics, Henan Normal University, Xinxiang 453007, P. R. China}

\emailAdd{fazlollah.hajkarim@ou.edu}
\emailAdd{g.a.white@soton.ac.uk}
\emailAdd{xiaoyangphy@gmail.com}

\abstract{
The gravitational wave spectrum arising from using the full velocity profile is well known to differ qualitatively from analytic fits to a broken power law. Former studies have shown that unlike the uncertainties arising from thermal field theory, more diligence in the hydrodynamics can sometimes have limited benefit. However, this was shown in the context of broken power law fits. We test the benefits of some recent calculations in modeling the spectrum, including new developments in adjustments of the low frequency tail to be consistent with causality, but we use the full velocity profile. We find the spectral shape information has a heightened sensitivity to the speed of sound which can be demonstrated analytically, however for our benchmark model this still results in a modest difference. The reason for a heightened sensitivity is because the velocity at the boundary is quite sensitive to the speed of sound, which in turn means a small change to the speed of sound can have a large change to the shape of the velocity profile. Furthermore, even modest changes in the product $\alpha \kappa$ can make non-trivial changes to the shape around the peak. Finally, there are many points where adjusting the infrared behavior to be consistent with causality is affecting the spectrum near its peak. All this implies that the spectrum is sensitive to five thermal parameters rather than four which gives hope that an observation of a gravitational wave spectrum from a first order cosmological phase transition could eventually give even more information about the underlying microphysics responsible.
}

\date{}

\begin{document}
\maketitle
\flushbottom

\section{Introduction}
\label{sec:intro}

Gravitational waves (GWs) discovery has opened the most novel probe of both astrophysical phenomena and the high‐energy frontier of the early universe \cite{LIGOScientific:2018mvr, LIGOScientific:2018jsj}. After the first detections of GW from merging black holes by LIGO, a new potential for probing cosmological sources of gravitational waves that is inaccessible to terrestrial experiments have been emerged \cite{LIGOScientific:2018mvr, LIGOScientific:2018jsj}. In particular, during a first order phase transition (FOPT) in the early Universe bubbles of a new phase nucleate in a metastable vacuum and expand then collide which can generate stochastic backgrounds of gravitational waves through bubble collisions, acoustic waves, and turbulence in the primordial plasma \cite{Mazumdar:2018dfl,Athron:2023xlk}. Detection or tight constraints on such signals would reveal details of particle interactions and the  thermal history  of universe at energy scales far beyond current collider reach \cite{Caprini:2015zlo,Caprini:2019egz, Weir:2017wfa,Schwaller:2015tja,Ellis:2018mja,Ellis:2020awk,Hindmarsh:2013xza, Hindmarsh:2015qta,Hindmarsh:2017gnf,Cutting:2018tjt}. 
Since the GW spectrum generated by a cosmological first order phase transition represents an approximate double broken power law, a possible detection signal can give some information about the origin of stochastic cosmological gravitational wave backgrounds \cite{Giese:2021dnw, Gowling:2022pzb, Boileau:2022ter,Guo:2024gmu}.

Grand unified theories and other beyond standard model physics scenarios can produce a sufficiently strong first‐order transitions at different scales that could produce observable gravitational waves \cite{Witten:1984rs, Hogan:1986qda, Kosowsky:1991ua, Turner:1990rc,  Chala:2016ykx, Huber:2008hg, Moore:2000jw, Steinhardt:1981ct,  Krauss:1991qu, Kamionkowski:1993fg, Grojean:2006bp, Enqvist:1992va, Joyce:1994fu,Dorsch:2016nrg, Konstandin:2010cd,Athron:2023xlk,Ghosh:2023aum, Baldes:2016rqn, Haba:2019qol,No:2011fi, Konstandin:2006nd, Hashino:2016rvx,Ashoorioon:2009nf, Bodeker:2009qy,Dine:1992wr, Apreda:2001us, Cai:2017tmh, Ananda:2006af, Boyle:2007zx, Binetruy:2012ze, Gleiser:1993hf, Gleiser:1998rw, Linde:1978px, Linde:1981zj,  Kosowsky:1992rz, Kosowsky:1992vn, Caprini:2007xq, He:2020ivk, Madge:2018gfl,Chang:2022tzj,Harigaya:2022ptp,Fujikura:2018duw,Dev:2016feu,Baldes:2018emh,Baldes:2021vyz,Azatov:2021ifm,Jinno:2016knw,Jinno:2017ixd,Lewicki:2021pgr, Croon:2018erz, Croon:2019ugf,Croon:2020cgk,Kakizaki:2015wua,Kobakhidze:2017mru, Chao:2017vrq}. Calculating the amplitude and spectral shape of these GW using key parameters such as bubble‐wall velocity, latent heat release, and nucleation rate is done in the literature and used for checking the detectability of GW signals with current and future observatories \cite{Mazumdar:2018dfl,Athron:2023xlk}. Ground based GW detectors due to their shorter arms can probe higher frequencies to identify characteristic of compact‐object mergers \cite{Moore:2014lga,KAGRA:2013rdx,Sathyaprakash:2012jk}.
The next generation of space‐based detectors like LISA and DECIGO can probe the millihertz to decihertz band that is the range for the electroweak scale phase transitions \cite{Moore:2014lga, Baker:2019nia, Cutler:1997ta, Klein:2015hvg, Allen:1997ad, Phinney:2001di, Caldwell:2018giq, Caldwell:2019vru,Barausse:2020rsu}. 
Moreover, pulsar timing arrays can probe nanohertz regime that is around the scale of QCD transition \cite{Witten:1984rs, Hogan:1986qda, Kosowsky:1991ua,  Chala:2016ykx, Maggiore:1999vm}. If during reheating epoch or non-standard cosmologies a phase transition happens then it can lead to the enhancement or suppression of the produced spectrum for GW from FOPT due to the change of temperature scaling and change of Hubble rate~\cite{Banik:2025olw,Xiao:2024rsj,Barenboim:2016mjm}.

Simulation of gravitational wave production in the early Universe depends on the equation of state (EOS) of the primordial plasma. In addition, it is important to investigate how the vacuum energy is distributed between bubble‐wall collisions, sound waves, and the effect of turbulence in the motion of background fluid \cite{Espinosa:2010hh,Huber:2008hg,Hindmarsh:2013xza,Hindmarsh:2015qta,Caprini:2009yp,Kosowsky:2001xp,Weir:2017wfa,Lewicki:2019gmv, Lewicki:2020jiv,Lewicki:2020azd}. The simplest case of FOPT can be considered using ``bag" that explains each phase as radiation‐like fluid with a constant vacuum‐energy shift. Thus it helps to check latent heat release and a large sound speed \cite{Espinosa:2010hh,Konstandin:2010cd}. There might be some deviations from the bag equation of state have been realized by recent lattice simulations of phase transitions \cite{Laine:2000rm,Gould:2019qek,Catumba:2025axv,Niemi:2024axp}. This can change the transition dynamics: the speed of sound may vary near the critical temperature, and friction on the bubble walls can prevent runaway acceleration \cite{Leitao:2014pda,Bodeker:2009qy,Megevand:2013hwa}, motivating ``beyond bag" models that include temperature dependent pressure and energy‐density functions with additional parameters \cite{Tenkanen:2022tly,Ai:2023see,Wang:2022lyd,Wang:2020nzm,Tian:2024ysd}.
There are other approximations beyond the bag model assumption that may include different aspects of a phase transition more precisely. The $\mu\nu$ model is one of them and is an extension of the bag model for a more detailed and realistic study of phase transition \cite{Giese:2020rtr,Giese:2020znk,Ai:2023see}. It has a different scaling parameter to pressure and energy density in the symmetric and broken phases instead of only having the bag constant in the equations of state.  This could lead to changes in effective equation of state parameter and speed of sound \cite{Wang:2022lyd,Giese:2020rtr,Giese:2020znk,Ai:2023see}. There are differences between the produced GW spectra between bag and beyond bag scenarios due to the change of equations of state that lead to different velocity and enthalpy profiles and consequently a distinguishable GW spectrum from FOPT 
\cite{Espinosa:2010hh,Wang:2021dwl,Wang:2022lyd,Giese:2020rtr,Wang:2020nzm}. 

Some fit functions based on the sound shell model are proposed for the sound waves component of GW from FOPT that include the properties of the model \cite{Hindmarsh:2019phv,Gowling:2021gcy}. There is also a new fit function that includes the double broken power-law behavior derived from sound shell model calculations~\cite{Guo:2024gmu}. This new fit includes a more precise treatment of sound shell model. 
Moreover, recent studies propose the low-frequency tail of the GW spectrum of FOPT in the sound shell model can scale more precisely as $k^9$ (instead of scaling like $k^3$) \cite{RoperPol:2023dzg, RoperPol:2023bqa, Sharma:2023mao,Cai:2019cdl}. This gives a more accurate treatment of the underlying integral for calculation of the GW spectrum.
This effect can provide a peak shape with a distinguishable observable signal in future observations.

The recent calculations of FOPT using bag model assumption do not suggest much of a change in the peak spectrum \cite{Hindmarsh:2019phv,Gowling:2021gcy}. This makes it naively questionable whether it is worth the diligence of going beyond the bag model. However, previous work has not used the full velocity profile which can qualitatively change the ``shape'' of the predicted power spectrum to resemble a double broken power law \cite{Guo:2021qcq,Guo:2024gmu,Hindmarsh:2019phv}. The peak amplitude in the broken power law analytic fit is only proportional to the average fluid velocity, which is not terribly sensitive to the speed of sound. By contrast, the shape of the spectrum depends on the velocity profile which is quite sensitive to the speed of sound. After making this case, we incorporate this into a picture of the highest diligence one can currently implement without performing a simulation.

This paper is organized as follows. First, we consider the impact of diligence factor by fixing a model. Then we consider the effect of low, moderate and high diligence over different frequencies in Sec.~\ref{sec:dilig}. Then we study the low frequency tail of GW from FOPT at low frequencies in Sec.~\ref{sec:ssm-gw}. In Sec.~\ref{sec:sec-sp-soun} we investigate the impact of the speed of sound on the spectrum of GW produced from FOPT. Finally, we summarize and conclude in section 
\ref{sec:summ-conc}.

\section{Diligence Over Different Frequencies}
\label{sec:dilig}

A cosmic first order phase transition can happen in the early universe when the Universe cools down due to the expansion and may experience a possible symmetry breaking. It ideally requires high powered simulations to study. At present, it is not practical to perform simulations to analyse entire models with many parameters \cite{Hindmarsh:2015qta,Cutting:2018tjt}. Thus, the community attempts to build the technology to capture as much of the physics of a first order phase transition as possible \cite{Mazumdar:2018dfl,Athron:2023xlk,Hindmarsh:2020hop}. This leads to different members of the community to enact different levels of diligence depending on how much it is worth adding extra physical effects. In this section, we summarize several common approaches and also motivate the need that in assessing the value of extra extra diligence, one needs to consider the change to the entire power spectrum and not just the peak.

First we review the physics of a cosmological first order phase transition. 
The thermal effective potential of the scalar field changes with temperature when the universe expands and cools down. At the critical temperature 
$T_c$, two degenerate minima appear in the scalar field potential. The transition of the scalar field between the symmetric vacuum and the broken one starts a first-order electroweak phase transition. The broken vacuum below $T_c$ becomes energetically favorable and the bubbles of the true vacuum may nucleate within the metastable symmetric phase through thermal tunneling~\cite{Linde:1983gd,Linde:1978px}. 
These bubbles then expand due to the vacuum energy difference and the  plasma friction. Then they collide and merge as the broken phase percolates through space. During this process, the  dynamics of bubble collisions, sound waves in the plasma, and subsequent turbulence can generate stochastic gravitational waves~\cite{Witten:1984rs, Hindmarsh:2013xza,Kamionkowski:1993fg}. It provides a window into the thermal history of the early universe that can be tested by experiments. The contribution from sound waves dominates at high temperatures~\cite{hogan1986gravitational,Hindmarsh:2013xza,Hindmarsh:2015qta}.  Consequently, we mainly focus on gravitational waves generated by sound waves. There are  multiple approaches for  computing the characteristic quantities of the phase transition. Each step of these calculations include choices that can lead to different final results. There are characteristic temperatures can be defined in different ways (e.g., $T_n$, $T_p$)~\cite{Athron:2023rfq,Athron:2023xlk}. The efficiency of energy transfer is estimated using fitting functions or derived from hydrodynamic models~\cite{Cai:2023guc,Wang:2022lyd,Si:2025vdt,Giese:2020znk,Espinosa:2010hh, Giese:2021dnw,Giese:2020rtr}. Moreover, the final gravitational wave spectrum can be obtained from lattice-based fits~\cite{Zhou:2020ojf,Xie:2020bkl,Athron:2023xlk,Caldwell:2022qsj}, model calculations such as the sound shell model~\cite{Cai:2023guc, Guo:2020grp,Guo:2023gwv,RoperPol:2023dzg, Sharma:2023mao}, or full numerical simulations~\cite{Hindmarsh:2013xza,Hindmarsh:2015qta,Hindmarsh:2017gnf,Hindmarsh:2019phv,Hindmarsh:2020hop,Caprini:2024gyk,Jinno:2022mie}. These differences represent various levels of approximation. To verify these effects, we compare gravitational wave spectra obtained under different levels of diligence.  This can clarify how computational assumptions can lead to observable predictions.

\subsection{Lowest Diligence}

In this paper we try to argue that the qualitative differences that arise from a more careful calculation are worth a model builders time. We consider the lowest diligence case here following the approach done in Ref.~\cite{Guo:2021qcq}. Additionally, the lowest diligence approach is a good framework to introduce the concepts behind the main thermal parameters that shape the gravitational wave power spectrum.
 
The nucleation rate per unit time per unit volume is calculated using an approximation for the path integral, 
\begin{equation}
  \Gamma = A T^4 e^{-\frac{S_3}{T}}. 
\end{equation}
Here $S_3$ is an extremum of the Euclidean action that can describe a bubble wall. We try to find it by minimizing the action given by \cite{Mazumdar:2018dfl,Athron:2023xlk}
\begin{equation}
  S_3 = 4\pi \int dr~r^2 \left[\frac{1}{2} \left(\frac{d \phi}{d r}\right)^2 + V_{\rm eff}(\phi; T)\right] .
\end{equation}
The pre-factor $A$ includes the loop corrections which are typically of order $\mathcal{O}(1)$. Using \texttt{BubbleDet}~\cite{Ekstedt:2023sqc} one can find the precise value of $A$. The number of bubbles within a Hubble volume reaches unity at the nucleation temperature $T_n$ as shown in the below 
\begin{equation}
  \int_{T_n}^{\infty} \frac{dT}{T} \frac{\Gamma(T)}{H(T)^4} \sim \mathcal{O}(1), 
\end{equation}
where $H(T)$ is the Hubble rate at the temperature $T$. Another quantitative representation of above condition in a radiation-dominated universe can be written as
\begin{equation}
  \frac{S_3}{T} \approx 140.
\end{equation}
The inverse duration time of the phase transition is defined by \cite{Mazumdar:2018dfl,Athron:2023xlk}
\begin{equation}
  \beta = H_n T_n \left.\frac{d S_3/T}{d T}\right|_{T= T_n}, 
\end{equation}
where $H_n$ is the Hubble rate corresponding to the nucleation temperature $T_n$. The ratio of change in the trace anomaly over the total energy density defines the transition strength. During the radiation domination epoch it can be defined as 
\begin{equation}
  \alpha = \left.\frac{\Delta V_{\rm eff} - \frac{T}{4}\frac{d \Delta V_{\rm eff}}{d T}}{\rho_{\rm rad}} \right|_{T =T_n}, 
\end{equation}
Here the difference between the false vacuum and the true vacuum is denoted by $\Delta$. The radiation energy density is shown by $\rho_{\rm rad}$. The peak of the gravitational wave spectrum depends on $\alpha$ and $\beta$.

The gravitational wave spectrum is approximated by a broken power law with just two observables, the peak frequency and amplitude \cite{Guo:2020grp,Mazumdar:2018dfl,Athron:2023xlk}
\begin{equation} 
\label{eq: GW fitting}
  \Omega_{\rm GW}^{\rm sw}h^2 = 8.5 \times 10^{-6} \left(\frac{100}{g_*}\right)^{\frac{1}{3}} \left(\frac{\kappa_{\rm sw} \alpha}{1 + \alpha}\right) \left(\frac{H_n}{\beta}\right)v_w S_{\rm sw}(f) \ .
\end{equation}
In the above, the number of degrees of freedom is $g_* \approx 100$ that is approximated at the nucleation temperature \cite{Drees:2015exa}. The parameter $\kappa_{\rm sw}$ is the efficiency factor that defines the fraction of the bulk kinetic energy in the plasma relative to the available vacuum energy. Bubble wall velocity is denoted by $v_w$.  The spectrum shape function $S_{\rm sw}$ can be obtained from lattice simulation or a specific model. In lowest diligence, we calculate the $\kappa_{\rm sw}$ by the fitting formula derived from the hydrodynamics based on the bag model~\cite{Espinosa:2010hh}. The spectral function $S_{\rm sw}$ is typically a broken power law when only the mean bubble separation is considered. However, since the gravitational wave spectrum also depends on the sound shell thickness, it is necessary to introduce a second characteristic scale. The spectral function follows a double broken power law shape. In our previous work, we obtained a new double broken power law by fitting the results of the sound shell model~\cite{Guo:2024gmu}
\begin{equation}
\label{eq:newfit}
S_{\rm sw}(f) = \left(\frac{f}{\tilde{s}_0}\right)^9 \cdot \frac{\left(2 + \tilde{r}_b^{-12 + \tilde{b}}\right)}{\left[\left(\frac{f}{\tilde{s}_0}\right)^{\tilde{a}} + \left(\frac{f}{\tilde{s}_0}\right)^{\tilde{b}} + \tilde{r}_b^{-12 + \tilde{b}} \cdot \left(\frac{f}{\tilde{s}_0}\right)^{12}\right]}\, \, , 
\end{equation}
where the parameters $\tilde{\Omega}_p$, ${\tilde s}_{0}$, $\tilde{a}$, $\tilde{b}$ and ${\tilde r}_b = f_b / f_p$ are calculated by the numerical solution of sound shell model and the computed fit function for the GW spectrum~\cite{Guo:2024gmu}. The ratio between the peaks is defined by $\tilde{r}_b$. The paramter $\tilde{b}$ specifies the spectral slope between the two frequency peaks.  The infrared behavior of the GW spectrum is controlled by the quantity $\tilde{a}$~\cite{Guo:2024gmu}. Since we intend to explore the differences in gravitational wave predictions under varying levels of diligence (including the fitting formula from our previous sound shell model) we adopt the double broken power law to ensure consistency.

\subsection{Moderate Diligence}
A more common approach, that tries to capture some of the most numerically significant corrections to the lowest diligence approach without getting too far into the weeds we refer to as the moderate diligence approach and was used in the recent LISA review \cite{Caprini:2019egz}. A more refined treatment of the characteristic temperature and the peak amplitude of the GW from FOPT is considered at the moderate diligence level. The nucleation temperature is not an ideal choice for a characteristic temperature. In cases of slow or strongly supercooled first-order phase transitions, the transition can still proceed even if the nucleation temperature is nonexistent~\cite{Athron:2022mmm, Wang:2020jrd, Kierkla:2023von}. The percolation temperature, $T_p$, defined as the temperature at which the false vacuum volume fraction drops to 70\%, closely approximates the nucleation temperature when it exists, while also providing a meaningful description in the aforementioned special cases. It thus serves as a robust and reliable alternative. The percolation temperature can be roughly estimated by the following equation \cite{Guo:2021qcq}
\begin{equation}
  \frac{S_3(T_p)}{T_p} = 131 - \mathrm{log}\left(\frac{A}{T_p^4}\right) - 4 \mathrm{log}\left(\frac{T_p}{100 {\rm GeV}}\right) - 4 \mathrm{log}\left(\frac{\beta(T_p)/H}{100}\right) + 3 \mathrm{log}(v_w), 
\end{equation}
where $\mathrm{log}\left(\frac{A}{T_p^4}\right) \approx 14$ for the electroweak phase transition. Interestingly, in the case of Standard Model Effective Field Theory (SMEFT), the above fit actually was worse than the lowest diligence fit \cite{Croon:2020cgk}. In general, during a supercooled phase transition, the existence of $T_p$ does not necessarily guarantee that the transition will complete. The most rigorous approach is still to compute the false vacuum fraction and verify whether it decreases to zero. With our choice of characteristic temperature, the strength factor $\alpha$ and inverse duration time $\beta$ become
\begin{align}
  \alpha &= \left.\frac{\Delta V_{\rm eff} - \frac{T}{4}\frac{d \Delta V_{\rm eff}}{d T}}{\rho_{\rm rad}} \right|_{T =T_p}, \notag \\
    \beta &= H_p T_p \left.\frac{d S_3/T}{d T}\right|_{T=T_p}.
\end{align}

In Eq.~\eqref{eq: GW fitting}, the source active time is overestimated and does not actually reach a full Hubble time. We can use characteristic quantities to roughly estimate the finite lifetime of sound waves, which leads to a suppression of the peak gravitational wave spectrum
\begin{align} \label{eq: sp1}
  \Omega_{\rm sw} &\to \Omega_{\rm sw} t_{\rm sw} H_p \notag, \\
  t_{\rm sw} &= {\rm min} \left[ \frac{1}{H_p}, \frac{R_{*}}{U_f} \right], 
\end{align}
where $R_{*}$ is the mean bubble separation, which can be related to $\beta$ as $R_{*} = (8\pi)^{1/3} v_w / \beta$ by expanding the action $S_3$ to first order, $U_f$ is the root-mean-square fluid velocity and can be obtained from hydrodynamic.

\subsection{High Diligence}
At the high diligence level, all variables appearing in the lattice-based gravitational wave formula will be computed with the highest possible accuracy, avoiding the use of fitting formulas or heuristic expressions whenever possible. In a previous paper, this was limited to the calculation of the peak amplitude and frequency using a broken power law fit. Here, we discuss the use of the full velocity power spectrum, rather than averaging, as in this context the use of the bag model can lead to a qualitatively different power spectrum, even if the peak amplitude prediction does not change that much. We of course do not claim that the highest diligence is the same level of accuracy as a simulation \cite{Hindmarsh:2013xza,Hindmarsh:2015qta, Hindmarsh:2017gnf,Cutting:2019zws,Gould:2021dpm}.

For the characteristic temperature, we first compute the false vacuum fraction, which we donate as $h$, in a precise \cite{Hindmarsh:2019phv} 
\begin{equation}
  h(t_c, t) = \mathrm{exp} \left[-\frac{4 \pi}{3} \int_{t_c} ^t d t' \Gamma(t') a^3(t') r(t', t)^3 \right], 
\end{equation}
where $r (t', t)$ is the comoving radius of a bubble nucleated at $t'$ and measure at $t$ 
\begin{equation}
  r(t', t) = \int_t'^t dt'' \frac{v_w}{a(t'')} = v_w(\eta' - \eta), 
\end{equation}
for FLRW space where $\eta$ is the conformal time. In a radiation-dominated universe, the measure can be changed through, 
\begin{equation}
  \frac{d T}{d t }= -HT.
\end{equation}
The phase transition is mature enough that the bulk of graviatational waves are being produced when the false vacuum fraction has been reduced to, 
\begin{equation}
  h(T_c, T_f) = \frac{1}{e}.
\end{equation}
In most cases, $T_p \approx T_f$ due to the short duration of the phase transition. However, for supercooled transitions these can differ and one must ensure that $T_f$ exists at all, the last one ends up with a Universe 
dominated by the false vacuum. 

For the bubble mean separation $R_{*}$, we track the evolution of the distribution of bubbles and use, 
\begin{equation}
  R_{*} = \left(\frac{V}{N_b}\right)^{1/3} = \left(\frac{1}{n_b}\right)^{1/3} \, , 
\end{equation}
where $n_b$ is the bubble density per proper volume and its evolution is determined by \cite{Hindmarsh:2019phv} 
\begin{equation} \label{eq: bubble density}
  \frac{d\left[n_b a^3(t)\right]}{d t} = \Gamma(t) h(t_c, t) a^3(t)
\end{equation}
with initial condition $n_b(t_c) = 0$. Rewriting Eq.~\eqref{eq: bubble density} in terms of conformal time $\eta$, we can express comoving $\beta_c$ as a function of $n_{b, c}$
\begin{equation}
  n_{b, c} = \frac{\beta_c^3}{8 \pi v_w^3}.
\end{equation}
Then $\beta$ can be expressed as 
\begin{equation}
  \beta(v_w) = \frac{a(\eta)}{a(\eta_f)} (8 \pi)^{1/3} \frac{v_w}{R_{*}(\eta)}\, , 
\end{equation}
where $\eta_f$ is the conformal time corresponding to $T_f$. Comparing with the previous $\beta$, the correction from cosmic expansion is now taken into account.

For the efficiency factor $\kappa_{\rm sw}$, we directly derive it from hydrodynamics. Specifically, we are interested in the free energy density of the 
scalar field-fluid system $\mathcal{F}(\phi, T)$. Once we obtain $\mathcal{F}$, the state variables of the system: the pressure density $p$ , the energy density $e$, the enthalpy density $w$ and the speed of sound $c_s$ can be expressed as \cite{Hindmarsh:2019phv}
\begin{align}
  p &= -\mathcal{F}, \notag \\
  e &= \mathcal{F} - T \frac{\partial \mathcal{F}}{\partial T}, \notag \\
  w & = p + e = - T \frac{\partial \mathcal{F}}{\partial T}, \notag \\
  c_s^2 &= \frac{\partial p}{\partial e}.
\end{align}
Using the above thermodynamic quantities, the energy-momentum tensor of the plasma is given by
\begin{equation}
  T^{\mu \nu} = w u^{\mu} u^{\nu} + g^{\mu \nu} p \, , 
\end{equation}
where $u^{\mu}$ is the four-velocity of the fluid and $g^{\mu \nu}$ is the inverse Minkowski metric. We plan to compute the hydrodynamic equations. We compute the projection of continuity equation $\partial_{\mu} T^{\mu \nu} = 0$. The parallel and perpendicular components w.r.t. to the fluid flow are calculated based on assuming  the self-similarity of solution~\cite{Giese:2020znk,Si:2025vdt}.  In other words, the appropriate solution only depends on $\xi = r/t$, where $r$ is the distance from the bubble center and $t$ is the time passed since the  nucleation of bubbles~\cite{Giese:2020znk,Si:2025vdt}. The hydrodynamic equations can then be written as
\begin{align}
  \frac{d v}{d \xi} &= \frac{2 v (1 - v^2)}{\xi (1 - v \xi)} \left(\frac{\mu(\xi, v)}{c_s^2} - 1\right)^{-1}, \notag \\
  \frac{d w}{d \xi} &=w \left(1 + \frac{1}{c_s^2}\right) \gamma^2 \mu(\xi, v) \frac{d v}{d \xi}, 
\end{align}
where $v(\xi)$ is the fluid velocity, $\gamma$ is the Lorenz factor and 
\begin{equation}
  \mu(\xi, v) = \frac{\xi -v}{1 - \xi v}.
\end{equation}
The direct numerical solution of  the equation reveals that for each $\xi$, there could be two corresponding $v$ values. Therefore, we need boundary conditions to ensure that our solution is a single-valued function. The boundary condition for $v$ is obtained by integrating the continuity equations in the wall frame across the bubble wall, 
\begin{align} \label{eq:match-condition}
  v_{+}v_{-} &= \frac{p_s(T_{+}) - p_b(T_{-})}{e_s(T_{+}) - e_b(T_{-})}\, , \notag \\
  \frac{v_{+}}{v_{-}} &= \frac{e_b(T_{-}) + p_s(T_{+})}{e_s(T_{+}) + p_b(T_{-})} \, , 
\end{align}
where the subscript $+$($-$) denotes the quantity in front (behind) of the bubble wall, while the subscript 
$s$ ($b$) represents quantity in the symmetric (broken) phase.

By imposing this boundary condition, the hydrodynamic equations can yield three types of solution: deflagration, hybrid, and detonation. The plasma is stationary in front of the bubble wall when the bubble moves. This creates a rarefaction wave behind it  known as a detonation~\cite{Giese:2020znk,Espinosa:2010hh}. 

For the deflagration case, the bubble wall velocity is below the speed of sound in the broken phase.
It can be identified by a shock wave in front of the bubble wall and a stationary plasma behind it. 
In a hybrid regime, the wall velocity is less than the Jouguet velocity and becomes larger than the speed of sound in the broken phase (see below)~\cite{Giese:2020znk,Espinosa:2010hh}.
In principle, the free energy density $\mathcal{F}(\phi, T)$ could be fully determined by the particle physics model. However, directly using the realistic free energy density to solve the hydrodynamic equations mentioned above is very challenging, involving dealing with the temperature dependent speed of sound $c_s$ and the boundary conditions that cannot be explicitly written. To reduce the difficulty of solving the equations, we can perform a high-temperature expansion of the free energy density and use this as the basis to construct a simplified model. We then map our realistic model onto this simplified model to obtain an approximate solution. There are two popular types of simplified models: the bag model and the $\mu \nu$ model \cite{Giese:2020znk,Espinosa:2010hh, Giese:2020rtr}.

In bag model, the pressure $p$ and energy density $e$ are simplified as the bag equation of state 
\begin{align} \label{eq: bag model}
  p_s &= \frac{1}{3}a_{+} T^4 - \epsilon, ~~~~p_b = \frac{1}{3}a_{-} T^4, \notag \\
  e_s &= a_{+}T^4 + \epsilon, ~~~~~~e_b = a_{-}T^4, 
\end{align}
where the temperature independent vacuum energy is denoted by $\epsilon$. The value of degrees of freedom in the symmetric and broken phase are connected to the parameters $a_{+}$ and $a_{-}$ \cite{Giese:2020znk}. In this simple model, the matching condition \cref{eq:match-condition} could be organized as \cite{Hindmarsh:2019phv} 
\begin{align}
  v_{+}v_{-} &= \frac{1 - (1 - 3 \alpha_{b, +}) r}{3 - 3(1 + \alpha_{b, +})r}, \notag \\
  \frac{v_{+}}{v_{-}} &= \frac{3 + (1 - 3 \alpha_{b, +}) r}{1 + 3(1 + \alpha_{b, +})r}, 
\end{align}
where 
\begin{align} 
  \alpha_{b, +} = \frac{4\epsilon}{3w_{+}}, ~~r = \frac{w_{+}}{w_{-}}.
\end{align}
Here, $\alpha_{b, +}$ is the ratio of the released vacuum energy to the enthalpy near the bubble wall when the phase transition happens. It also represents the strength of the phase transition. Since the temperature near the bubble wall does not exactly match with the background temperature, the value of $w_{+}$
is generally unknown without explicit calculation. For convenience, we define $\alpha_b$ 
\begin{align} \label{eq: alpha bag}
  \alpha_{b} = \frac{4\epsilon}{3w_{s}}, 
\end{align}
to be an input parameter in the sound shell model instead of $\alpha_{b, +}$. Once the temperature profile is known, $\alpha_b$ and $\alpha_{b, +}$ can be converted into each other.

The speed of sound could be easily obtained and $c_{s, b}^2 = c_{s, s}^2 = 1/3$. With this assumption, the corresponding Jouguet velocity is given by
\begin{equation}
  \xi_J = \frac{\sqrt{\alpha_{b, +}(2 + 3 \alpha_{b, +})} + 1}{\sqrt{3}(1 + \alpha_{b, +})} .
\end{equation}

The $\mu \nu$ model incorporates deviations of the square of the speed of sound $c_s^2$ from $1/3$, but it also assumes that the temperature difference between inside and outside the bubble is minimal, i.e, $T_{+} \approx T_{-}$. The pressure density $p$ and $e$ of this model are defined as~\cite{Giese:2020znk}
\begin{align} \label{eq: mu_nu model}
  p_s &= \frac{1}{3}a_{+} T^{\mu} - \epsilon, ~~~~p_b = \frac{1}{3}a_{-} T^{\nu}, \notag \\
  e_s &= \frac{1}{3}a_{+}(\mu-1)T^{\mu} + \epsilon, ~~~~~~e_b = \frac{1}{3}a_{-}(\nu-1)T^{\nu}, 
\end{align}
where $\epsilon$ is still the temperature-independent vacuum energy as in bag model and we have 
 \begin{equation}
  \mu = 1 + \frac{1}{c_{s, s}^2}, ~~~~~\nu = 1 + \frac{1}{c_{s, b}^2}.
\end{equation}
The matching condition \cref{eq:match-condition} will be more complex and can be expressed as 
\begin{equation} \label{eq: mu nu boundary condition}
  \frac{v_{+}}{v_{-}} = \frac{\left(v_{+}v_{-}/c_{s, b}^2-1\right) + 3 \alpha_{\theta, +}}{\left(v_{+}v_{-}/c_{s, b}^2-1\right) + 3 v_{+}v_{-} \alpha_{\theta, +}}, 
\end{equation}
where
\begin{equation}
  \alpha_{\theta, +} = \frac{\theta_s - \theta_b}{3 w_{+}}, ~~{\rm with}~\theta_{s, b} = e_{s, b} -\frac{p_{s, b}}{c_{s, b}^2}.
\end{equation}

Here we introduce an improved definition of the phase transition strength, denoted as $\alpha_{\theta, +}$. Similarly as in the bag model, we usually use $\alpha_{\theta}$~\cite{Giese:2020rtr,Giese:2020znk}
\begin{equation}
\label{eq:alpha-theta}
  \alpha_{\theta} = \frac{\theta_s - \theta_b}{3 w_{s}}, 
\end{equation}
where this is an input parameter in the sound shell model. It generalizes the original parameter $\alpha_b$ defined in \cref{eq: alpha bag}. This includes additional model-dependent information and provides a more realistic explanation for particle physics models. 
If we assume $c_{s, s}^2 = c_{s, b}^2 = 1/3$, then $\mu\nu$ model becomes identical to the bag model and we obtain $\alpha_{\theta} = \alpha_{b}$. 
With the definition of $\alpha_{\theta}$, the Jouguet velocity is then modified as
\begin{equation}
\xi_J = \frac{\sqrt{3 \alpha_{\theta, +}\left(1 - c_{s, b}^2 + 3 c_{s, b}^2 \alpha_{\theta, +}\right)} + 1}{1/c_{s, b} + 3 c_{s, b}\alpha_{\theta, +}} .
\end{equation}

The speed of sound rarely deviates very far from $1/\sqrt{3}$ in realistic models \cite{Giese:2020znk}, so one might assume that there is little difference in the predictions of the bag model and $\mu \nu$ model, even if one uses the full velocity profile. However, it turns out that the boundary terms $v_\pm$ are surprisingly sensitive to small deviations in the speed of sound. To illustrate this, let us approximate the result \cref{eq: mu nu boundary condition} based on the following expansion 
\begin{eqnarray}
\\[6pt]
\frac{v_{+}v_{-}}{c_{s, b}^{2}}
&=&
3\, v_{+}v_{-}-9\, v_{+}v_{-}\, q+\mathcal{O}(q^{2}), 
\end{eqnarray}
\begin{eqnarray}
\frac{v_{+}}{v_{-}}
= 
\frac{3v_{+}v_{-}-1+3\alpha_{\theta, +}}
   {3v_{+}v_{-}-1+3v_{+}v_{-}\alpha_{\theta, +}}
\;+\;
q\, \frac{27\, v_{+}v_{-}\, \alpha_{\theta, +}\, (1-v_{+}v_{-})}
    {\left(-1+3v_{+}v_{-}+3v_{+}v_{-}\alpha_{\theta, +}\right)^{2}}
\;+\;\mathcal{O}(q^{2}).
\end{eqnarray}
The prefactor in the second term of above equation is a large number $27$ that will lead to a significant change in the profiles. 
As an example, assuming $v_{+} = 0.5$ and $v_{-} = 0.5$ and $\alpha_{\theta, +} = 0.1$ then one can obtain $v_{+}/v_{-} = - 0.28 + 16.53 ~q$ up to $\mathcal{O}(q^{2})$ where the second term is large even for a small value of $q$. Then a small change in the speed of sound of a broken phase can change the velocity profiles significantly. This is visible in the velocity profiles in the left panels of Fig.~\ref{fig: munv_bag_cs}. 
Based on above explanations as it is visible in the figures a small deviation in the speed of sound in the broken phase will lead to a distinguishable change in the velocity and enthalpy profiles and produced GW spectrum.

 We can also write the following ratio based on speed of sounds in symmetric and broken phases, the bag constant and the phase transition strength
\begin{equation}
\frac{w_{+}}{w_{-}} =
\frac{(1+c_{s,s}^{2})(1-c_{s,b}^{2})}
{(1+c_{s,b}^{2})\left[(1-c_{s,s}^{2})+6\,\alpha_{\theta,+}\,c_{s,s}^{2}\right]} .
\label{eq:w-ratio-final}
\end{equation}
Assuming the following condition as we did for the ratio of velocities 
\begin{eqnarray}
  c_{s, b}^{2}&=&\frac{1}{3}+q, \ \ |q|\ll 1\,.
\end{eqnarray}
We get the following fraction 
\begin{eqnarray}
\frac{w_+}{w_-}
&=&
\frac{1 + c_{s,s}^{2}}{2\left[(1 - c_{s,s}^{2}) + 6\,\alpha_{\theta,+}\,c_{s,s}^{2}\right]}
\left(1 - \frac{9}{4}\,q\right)
+ \mathcal{O}(q^{2}),
\end{eqnarray}
where we assume $\alpha_{\theta,+} = 0.1$ and $c_{s,s}^{2} = 1/3$ then we obtain this fraction
$w_+/w_- 
\simeq 0.769\,(1 - 2.25\,q)
+ \mathcal{O}(q^{2})$. Since the coefficient of $q$ is larger with a small change of $q$ and we have large change in the enthalpy profiles inside and outside of the bubble and gives a visible impact of $\mu \nu$ model.

For both model, the kinetic energy fraction $K$ is defined as \cite{Hindmarsh:2019phv} 
\begin{align}
  K &= \frac{\rho_{fl}}{e_s}, \notag \\
  \rho_{fl} &= \frac{3}{v_w^3} \int d \xi \xi^2 v ^2 \gamma^2 w, 
\end{align}
where $\rho_{fl}$ is the fluid’s kinetic energy. By definition, the efficiency factor $\kappa_{\rm sw}$ can be related to $K$ via
\begin{equation}
  K = \kappa_{\rm sw} \left(\frac{\theta_s - \theta_b}{4 e_s}\right), 
\end{equation}
leading to 
\begin{equation}
  \kappa_{\rm sw} = \frac{4 \rho_{fl}}{3 \alpha_{\theta} w_s}.
\end{equation}
The 
root-mean-square of velocity in the fluid around a single bubble $U_f$ can be computed from the fraction of kinetic energy \cite{Hindmarsh:2019phv} 
\begin{equation}
  U_f^2 = \frac{e_s}{w_s} K.
\end{equation}
To calculate the $\kappa_{\rm sw}$ more precisely, we adopted the $\mu \nu$ model and use the publicly available code in Ref.~\cite{Giese:2020znk} to numerically compute the kinetic energy efficiency for a given set of $c_s^2$ and $\alpha_{\theta}$.

For the peak value of GW spectrum, we adopt the analytically derived suppression factor due to finite lifetime of the source, which relies solely on the assumption that the sound waves are approximately stationary \cite{Gowling:2021gcy} 
\begin{equation} \label{eq: sp2}
  \Upsilon = \frac{1}{\sqrt{1 - 2 \tau_{\rm sw} H}}, 
\end{equation}
where $\tau_{\rm sw} = R_{*} / U_f$ is the active time of the source. In addition, the suppression effect from the finite lifetime of the source and numerical simulations show that the formation of reheated parts  of the metastable phase can reduce the speed of bubble walls and reheat the surrounding regions~\cite{Guo:2021qcq}. This introduces an additional suppression to the  peak of gravitational wave spectrum. To incorporate this effect, we extract the fluid velocity following Ref.~\cite{Guo:2021qcq} and define the ratio
\begin{equation} \label{eq: sp3}
  B = \frac{U_f}{U_{f}^{\text{extract}}}
\end{equation}
to account for the deviation. After taking into consideration all the factors, the final peak value of the GW spectrum is 
\begin{equation}
  \Omega_{\rm GW} \to \Omega_{\rm GW} \Upsilon B.
\end{equation}


\section{Sound Shell Model and GW Spectrum at Low Frequencies}
\label{sec:ssm-gw}
Beyond the use of lattice-based fitting formulas, gravitational wave production can also be modeled analytically or numerically. The comparison between these models and fit functions is important in calculating the uncertainties in gravitational wave spectrum.  The sound shell model gives an analytical framework to explain the dynamics of these acoustic waves \cite{Hindmarsh:2019phv}. Considering in an expanding universe with metric given by
\begin{equation}
  ds^2 = a^2(\eta)\left[ -d \eta^2 + (\delta_{ij} + l_{ij}) dx^i dx^j \right], 
\end{equation}
the time evolution of GW during radiation-dominated universe can be described by the following equation
\begin{equation}
  ( \partial_{\eta}^2 + k^2) h_{ij}(\eta, \mathbf{k} ) = \frac{ 6 \mathcal{H}_{*} \Pi_{ij}(\eta, \mathbf{k})}{\eta}, 
\end{equation}
where $h_{ij} = a l_{ij}$, $\mathcal{H}$ is the conformal Hubble constant. The anisotropic stress tensor has a transverse-traceless part that is shown by $\Pi_{ij}$. We follow Ref.~\cite{RoperPol:2023dzg} for the formalism we use here. If the source is active during $\eta_{*} < \eta < \eta_{\rm fin}$, we can use the initial condition $h_{ij}(\eta_{*}, \mathbf{k}) = h_{ij}^{'}(\eta_{*}, \mathbf{k}) = 0$ and the Green's function to obtain the below solutions \cite{RoperPol:2023dzg}
\begin{eqnarray}
h_{ij}(\eta, \mathbf{k}) &=&
\left\{
\begin{array}{ll}
\displaystyle \frac{6\, \mathcal H_*}{k}\int_{\eta_*}^{\eta} d\eta_1\, 
\frac{\Pi_{ij}(\eta_1, \mathbf{k})}{\eta_1}\, 
\sin\!\big[k(\eta-\eta_1)\big], & \eta_* \le \eta \le \eta_{\rm fin}, \\[1.0ex]
\displaystyle \frac{6\, \mathcal H_*}{k}\int_{\eta_*}^{\eta_{\rm fin}} d\eta_1\, 
\frac{\Pi_{ij}(\eta_1, \mathbf{k})}{\eta_1}\, 
\sin\!\big[k(\eta-\eta_1)\big], & \eta \ge \eta_{\rm fin}.
\end{array}
\right.
\label{eq:hij}
\end{eqnarray}
With this solution, the energy density of gravitational waves is defined as
\begin{equation}
  \rho_{\rm gw} = \frac{1}{32 \pi G a^2} \left<h_{ij}^{'} h_{ij}^{' *}\right>\, , 
\end{equation}
where $'$ denotes the derivative with respect to conformal time, then by the definition of the power spectrum of GW, we can obtain the below expressions if $k$ is much larger than the inverse of the conformal time at today  \cite{RoperPol:2023dzg}
\begin{equation}
  \Omega_{\rm GW}(k) = \frac{1}{\bar{\rho}} \frac{d \rho_{\rm GW}}{d {\rm ln} k} = \frac{3 k }{2} \mathcal{T}_{\rm GW} \int_{\eta_{*}}^{\eta_{\rm fin}} \frac{d \eta_1}{\eta_1} \int_{\eta_{*}}^{\eta_{\rm fin}} \frac{d \eta_2}{\eta_2} E_{\Pi}(\eta_1, \eta_2, k) {\rm cos}\left[k (\eta_1 - \eta_2)\right], 
\end{equation}
where $\mathcal{T}_{\rm GW}$ is the red-shift factor
\begin{equation}
  h^2 \mathcal{T}_{\rm GW} = 1.6 \times 10^{-5} \left(\frac{100}{g_*}\right)^{\frac{1}{3}}\,,
\end{equation}
and $E_{\Pi}$ is the unequal time correlator (UETC) of the shear stress \cite{RoperPol:2023dzg}
\begin{equation}
  \left<\Pi_{ij}(\eta_1, \mathbf{k}_1) \Pi_{ij}^{*}(\eta_2, \mathbf{k}_2)\right> = (2 \pi)^6 \delta^3(\mathbf{k} - \mathbf{k}_2)\frac{E_{\Pi}(\eta_1, \eta_2, k)}{4 \pi k^2}.
\end{equation}
Consequently, the key to calculate the GW spectrum lies in obtaining the UETC of the shear stress, relating to the energy-momentum tensor $T_{ij}$. For sound waves (or plasma), its energy-momentum tensor at first order is 
\begin{equation}
  T_{ij} = \bar{w} u_i u_j + p \delta_{ij} \, , 
\end{equation}
where $\bar{w}$ is the averaged enthalpy, $\gamma$ is the Lorentz factor and $u_i = \gamma v_i$ is the spatial components of the four-velocity of plasma. In addition, at first order, $\gamma \sim 1$. The UETC of the shear stress suggests that we need the UETC of the energy-momentum tensor, which involves the four-point correlation function of the velocity field, $\langle u_i u_j u_l u_k \rangle$. By assuming the fluid velocity field to be Gaussian and applying Wick's theorem, the four-point correlation function of the velocity components can be reduced to a linear combination of products of two-point correlation functions. We define the two-point correlation function of the velocity field in Fourier space as \cite{RoperPol:2023dzg}
\begin{equation}
\left<u_i(\eta_1, \mathbf{k}), u_j(\eta_2, \mathbf{k}_2)\right> = (2 \pi)^6\hat{k}_i \hat{k}_j\delta^3(\mathbf{k} - \mathbf{k}2)\frac{2 E_{\rm kin}(\eta_1, \eta_2, k)}{4 \pi k^2}, 
\end{equation}
where $E_{\rm kin}(\eta_1, \eta_2, k)$ represents the unequal-time kinetic energy spectrum. Once this correlation function is known, the UETC of the shear stress can be constructed as \cite{RoperPol:2023dzg}
\begin{align}
  E_{\Pi}(\eta_1, \eta_2, k) &= 2 k^2 \bar{w}^2 \int_{-1}^1dz \int_0 ^{\infty}dp \frac{p^2}{\tilde{p}} (1-z^2)^2 \notag \\
  & \times E_{\rm kin}(\eta_1, \eta_2, p) E_{\rm kin}(\eta_1, \eta_2, \tilde{p}), 
\end{align}
where $z = \hat{\mathbf{k}} \cdot \hat{\mathbf{p}}$ and $\tilde{\mathbf{p}} = \mathbf{k} - \mathbf{p}$.

By the conservation of energy and momentum, $\partial_{\mu} T^{\mu \nu} = 0$ and assuming radial symmetry around the bubble nucleation site, the evolution of the fluid field $u_i$ can be determined as \cite{Hindmarsh:2019phv}
\begin{align}
  \lambda^{'}(\eta, \mathbf{k}) - i k_i u_i(\eta, \mathbf{k}) &= 0, \notag\\
  u_i^{'}(\eta, \mathbf{k}) - i k_i c_s^2 \lambda(\eta, \mathbf{k}) &= 0, \notag\\
\end{align}
where $\lambda = (e - \bar{e}) / \bar{w}$ is the normalized energy fluctuations. Its solution is longitudinal velocity field
\begin{equation}
  u_i = \hat{k}_i u = \hat{k}_i \sum_{s = \pm} A_s(\mathbf{k}) e^{isw (\eta - \eta_{*})}.
\end{equation}
To determine the coefficients $A_{\pm}$, the sound shell model assumes the whole velocity field is the superposition of each velocity field surrounding the corresponding bubble, thus 
\begin{equation}
  A_{\rm}(\mathbf{k}) = \sum_{n =1}^{N_b} \mathcal{A}_{\pm} \tilde{T}_{\rm n}^{3} e^{i \mathbf{k} \cdot \mathbf{x}_{0}}, 
\end{equation}
where, $\tilde{T}_{\rm n}$ is the lifetime of the $n$-th bubble and $x_0$ is its nucleation location. The amplitude functions $\mathcal{A}_{\pm}(\chi = k \tilde{T})$ are \cite{Hindmarsh:2019phv}
\begin{align}
  \mathcal{A}_{\pm}(\chi) &= \frac{-i}{2}\left[f^{'}(\chi) \pm i c_s l(\chi) \right], \notag \\
  f(\chi) &= \frac{4 \pi}{\chi}\int_0 ^{\infty} d \xi v_{ip}(\xi) \mathrm{sin}(\chi \xi), \notag \\
  l(\chi) &= \frac{4 \pi}{\chi}\int_0 ^{\infty} d \xi \lambda_{ip}(\xi) \mathrm{sin}(\chi \xi), 
\end{align}
where $v_{ip}$ and $\lambda_{ip}$ represent the velocity profile and energy fluctuation profile of a single bubble, respectively and can be calculated via bag model or $\mu \nu$ model mentioned above. Once we obtain the longitudinal velocity profile, we can calculate the $E_{\rm kin}$ via the bubble lifetime distribution $\nu(\tilde{T})$
\begin{equation}
  E_{\rm kin}(k) = \frac{k^2}{2 \pi^2 \beta^6 R_{*}^3} \int_0^{\infty} d\tilde{T} \nu(\tilde{T}) \tilde{T}^6 \frac{1}{4}\left[f^{'2}(\chi) + c_s^2 l^2(\chi)\right]
\end{equation}
and express the final spectrum in terms of $E_{\rm kin}$ \cite{RoperPol:2023dzg}
\begin{align}
  \Omega_{\rm GW}(\delta \eta, k) &= 3 \bar{w}^2 k^3 \mathcal{T}_{\rm GW} \int_{-1}^{1} (1 - z^2)^2 dz \int_0 ^{\infty} dp \frac{p^2}{\tilde{p}^4} E_{\rm kin}(p) E_{\rm kin}(\tilde{p}) \Delta(\delta \eta, k, p, \tilde{p}) \,, \notag \\
  \Delta(\delta \eta, k, p, \tilde{p}) &= \int_{\eta_{*}} ^{\eta_{\rm fin}} \frac{d \eta_1}{\eta_1} \int_{\eta_{*}}^{\eta_{\rm fin}} \frac{d \eta_2}{\eta_2} \mathrm{cos}\left[p c_s (\eta_{2} - \eta_{1})\right] \mathrm{cos}\left[\tilde{p} c_s (\eta_{2} - \eta_{1})\right] \mathrm{cos}\left[k c_s (\eta_{2} - \eta_{1})\right] \, , 
\end{align}
where $\delta \eta$ is the source active time. The function $\Delta$ can be calculated explicitly using the cosine and sine integral functions, ${\rm Ci}(x)$ and ${\rm Si}(x)$. As pointed out in Ref.~\cite{RoperPol:2023dzg, Sharma:2023mao}, in the original formulation of the sound shell model, $\Delta$ was approximated by a Dirac delta function, which leads to an unphysical $k^9$ behavior in the low-frequency regime. A more careful evaluation, however, restores the physically consistent $k^3$ scaling, as required by causality.

\subsection{Example of a BSM Model: xSM Model}
We use the Standard Model singlet extension (xSM) as a benchmark model, whose tree-level effective potential is given by \cite{Vaskonen:2016yiu,Curtin:2014jma}
\begin{align}
  V_0 = - \mu H^{\dagger}H + \lambda (H^{\dagger}H )^2 + \frac{1}{2}\mu_s^2 S^2 + \lambda_{hs}H^{\dagger}HS^2 + \frac{1}{4}\lambda_S S^4.
\end{align}
Following Ref.~\cite{Curtin:2014jma}, in this work, we focus on adding one real singlet with a mass larger than $m_h$ to avoid exotic higgs decays, and an unbroken $Z_2$ symmetry under which $S$ avoid singlet-higgs mixing. This constraint ensures that the additional scalar field does not acquire a vacuum expectation value at any temperature, thereby reducing the dependence of physical quantities on the model parameters and allowing us to more clearly identify how these physical quantities vary with the parameters. 

Under this constraint, the full finite-temperature effective potential can be written as
\begin{eqnarray}
V_{\rm eff}(h,T) &=& V_0(h) + V^{\rm CW}_0(h) + V_T(h,T),
\end{eqnarray}
where $V_0$ denotes the tree-level Higgs potential, the term $V^{\rm CW}_0$ is the zero-temperature one-loop Coleman-Weinberg correction under on-shell renormalization scheme \cite{Vaskonen:2016yiu,Curtin:2014jma}
\begin{eqnarray}
V^{\rm CW}_0 &=& 
\sum_i (-1)^{F_i} \frac{g_i}{64\pi^2}
\left[
m_i^4(h) \left( \log \frac{m_i^2(h)}{m_i^2(v)} - \frac{3}{2} \right)
+ 2 m_i^2(h) m_i^2(v)
\right],
\end{eqnarray}
and the one-loop finite-temperature contribution to the effective potential takes the form
\begin{eqnarray}
V_T(h,T) &=&
\sum_i (-1)^{F_i} \frac{g_i T}{2\pi^2}
\int dk\, k^2\,
\log\left[
1 - (-1)^{F_i}
\exp\left(
\frac{1}{T}\sqrt{k^2 + M_i^2(h)}
\right)
\right].
\end{eqnarray}
The fermion number $F_i$ equals 1 for fermions and 0 for bosons. $g_i$ is the number of degrees of freedom in the relativistic regime for particle species $i$. $m_i(h)$ is the field depended mass and can be expressed as \cite{Vaskonen:2016yiu,Curtin:2014jma}
\begin{align}
  m_t^2 &= \frac{\lambda_t}{2} h^2, \notag\\
  m_w^2 &= \frac{g^2}{4} h^2, \notag\\\
  m_w^2 &= \frac{g^2 + g'^2}{4} h^2, \notag\\\
  m_h^2 &= -\mu^2 + 3 \lambda h^2, \notag\\\
  m_s^2 &= -\mu_s^2 + \lambda_{hs}^2. 
\end{align}
In practice, we also neglect the numerically insignificant Goldstone contributions, as treating them properly near $h\sim 246$ GeV requires special care~\cite{Martin:2014bca}.

\subsection{FOPT GW Spectrum at Low Frequencies}

To ensure that the additional scalar field never acquires a vacuum expectation value during the phase transition, we require
\begin{equation}
  \mu_s^2 = m_s^2 - \lambda_{hs} h^2
\end{equation}
to remain strictly positive at all temperatures~\cite{Curtin:2014jma}. 
This condition generally favors a relatively large $m_s$ and a small $\lambda_{hs}$. 
However, for small $\lambda_{hs}$ and strictly positive $\mu_s^2$, it becomes increasingly difficult to find regions of parameter space that support a first-order phase transition. 
Therefore, in practice we restrict our scan to
\begin{equation}
  m_s \in [570~\mathrm{GeV},\, 630~\mathrm{GeV}], 
  \qquad
  \lambda_{hs} = 5.
\end{equation}
It is worth noting that Ref.~\cite{Curtin:2014jma} pointed out that in this parameter region the electroweak phase transition is largely driven by loop corrections. Thus, higher loop contributions to the effective potential may become important and require additional care. Since the primary goal of this work is not the precise computation of the effective potential itself, we confine our analysis to this simplified parameter region.

\begin{figure}[H]
  \centering
  \includegraphics[width=0.49\textwidth]{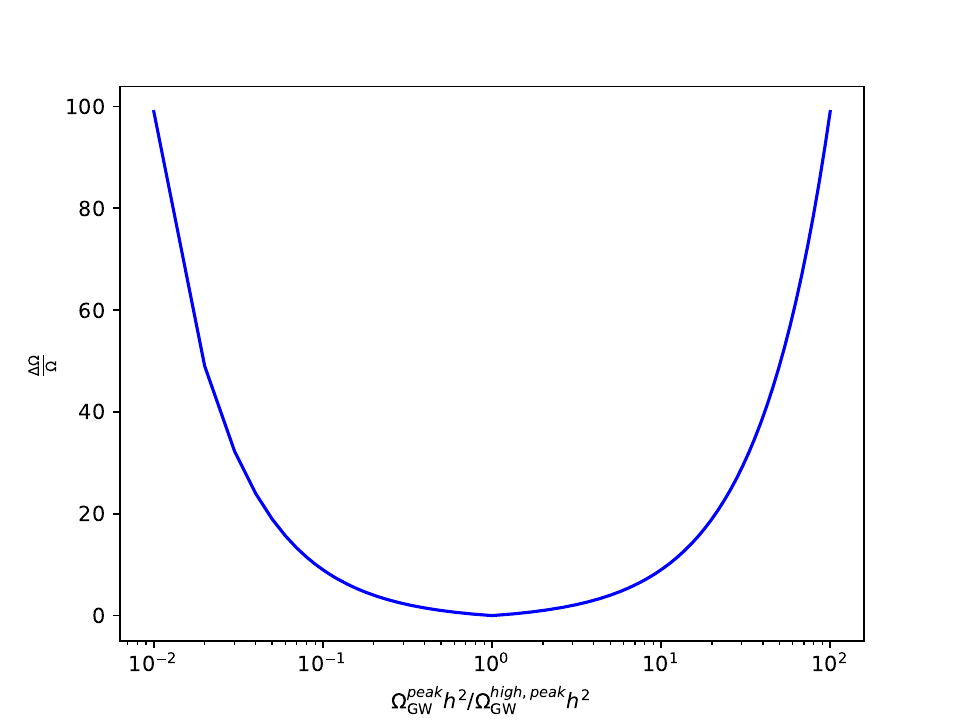}
  \includegraphics[width=0.49\textwidth]{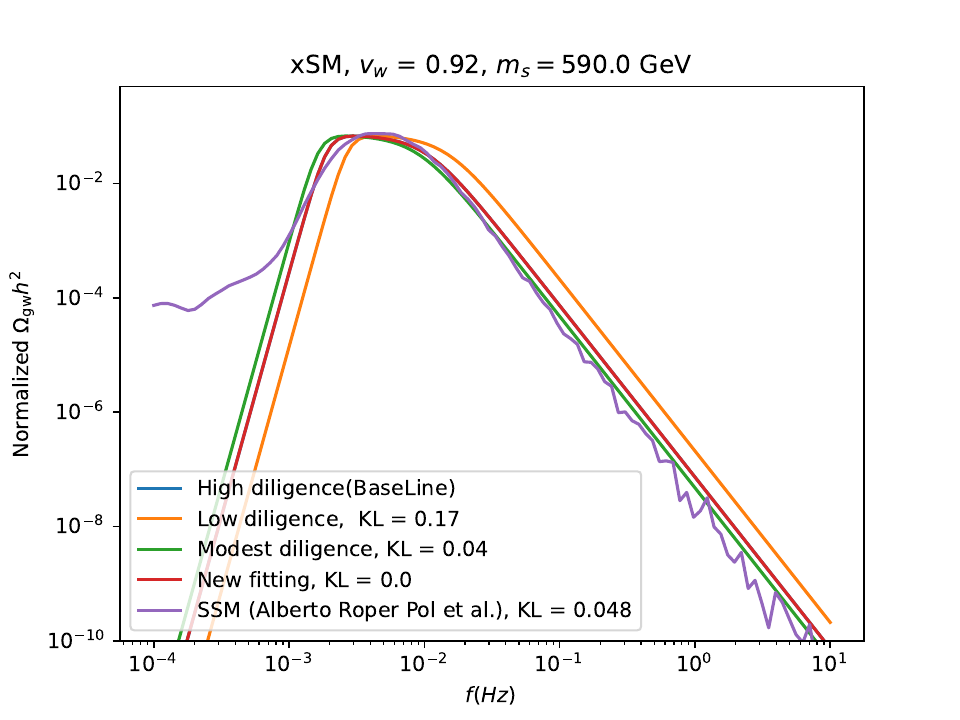}
\caption{Interpretation of the metrics used to quantify the differences between gravitational wave spectra. The left panel shows the relationship between the ratio of the target spectrum’s peak amplitude to the high-diligence result and the $\frac{\Delta \Omega}{\Omega}$. The right panel compares the normalized gravitational wave spectra for the chosen parameter set, illustrating cases with different values of the KL divergence. 
}
\label{fig: Delta and KL}
\end{figure}

To compare the gravitational wave spectra obtained under different levels of diligence, we take the high-diligence result as the baseline and define the following variable to estimate the uncertainty in the peak amplitude of the spectrum
\begin{equation}
  \frac{\Delta \Omega}{\Omega} = \frac{|\Omega^{j, \rm peak}h^2 - \Omega^{\rm high, peak}h^2|}{{\rm min}(\Omega^{j,\rm peak}h^2, \Omega^{\rm high,peak}h^2)}, 
\end{equation}
where $j = ({\rm low}, {\rm modest}, {\rm new fiting}, {\rm SSM })$. 
To provide an intuitive understanding, we plot the correlation between 
$\frac{\Delta \Omega}{\Omega}$ and the ratio of gravitational wave peak amplitudes, see Fig.~\ref{fig: Delta and KL} (Left). As shown, regardless of how much the target spectrum’s peak deviates from the baseline, the value of 
$\frac{\Delta \Omega}{\Omega}$ consistently increases. Therefore, a larger 
$\frac{\Delta \Omega}{\Omega}$ corresponds to a higher peak amplitude in the gravitational wave spectrum. Beyond the peak amplitude, the shape of the spectrum should also be included when comparing different GW spectra. Mathematically, the parameter $\frac{\Delta \Omega}{\Omega}$
only captures information about the peak amplitude, providing no insight into differences in peak frequency, or the infrared and ultraviolet behavior of the spectra. To quantify how the overall spectral distribution of a target spectrum deviates from the reference, we normalize each spectrum by the sum of all the corresponding sampling points. This removes information related to the peak itself and preserves only the shape information. In the ideal case where two spectra differ merely by a constant factor, this normalization yields identical distributions. Based on this reason, we introduce the Kullback–Leibler (KL) divergence to quantitatively compare differences in the spectral shapes \cite{Kullback:1951zyt,rezende2015variational,csiszar2004information,edwards2008elements}
\begin{equation}
  {\rm KL~ Divergence}(\Omega^j h^2, \Omega^{\rm high}h^2) = \sum_i \Omega^j_{i,\rm nor} h^2 {\rm log}(\frac{\Omega^j_{i, \rm nor} h^2}{\Omega^{\rm high}_{i,\rm nor} h^2})\, , 
\end{equation}
where $i$ is the index of sampling point and $j$ is the same as the above. It is straightforward to see that if the normalized GW spectra are identical, the corresponding KL divergence equals zero. 
Furthermore, when two spectra differ only by a global positive factor, and each spectrum is first normalized to unit total, their KL divergence is zero. Let
\begin{equation}
p_i=\frac{\Omega^{\,j}_i}{\sum_{k}\Omega^{\,j}_k},\qquad
q_i=\frac{\Omega^{\,\mathrm{high}}_i}{\sum_{k}\Omega^{\,\mathrm{high}}_k},
\end{equation}
then the KL divergence is \cite{Kullback:1951zyt,rezende2015variational}
\begin{equation}
\mathrm{KL}(p\|q)=\sum_i p_i \log\!\frac{p_i}{q_i}.
\end{equation}
If there is a constant $c>0$ such that $\Omega^{\,j}_i=c\,\Omega^{\,\mathrm{high}}_i$ for all $i$,
then we have $p_i=q_i$ and every log-ratio vanishes. This gives $\mathrm{KL}(p\|q)=0$.
In other words, with this normalization KL is \emph{scale-invariant} and measures \emph{shape} differences only. By contrast, a fractional metric such as $\Delta\Omega/\Omega$ captures \emph{amplitude} (i.e., overall scale) differences. \par 
The greater the discrepancy between the normalized spectra, the larger the KL divergence becomes. For an intuitive illustration of this behavior, several representative examples are shown in Fig. \ref{fig: Delta and KL} (Right). We observe that, when taking the high-diligence result as the reference, the normalized spectrum from the new fitting (red line) almost perfectly overlaps with it, resulting in a very small KL divergence that is close to  zero. Since the modest-diligence result (green line) also shows only a minor deviation from the high-diligence case, its KL divergence remains similarly small. However, the SSM result in Ref.~\cite{RoperPol:2023dzg} (purple line) identifies a larger deviation from the high-diligence spectrum in the infrared region.  This gives a smaller discrepancy in the ultraviolet region. Consequently, its KL divergence is larger than that of the modest-diligence case (orange line), but smaller than that of the low-diligence case, consistent with the fact that its deviation from the high-diligence spectrum is moderate across the full frequency range. Therefore, a larger KL divergence represents a greater difference in the distribution of the gravitational wave spectra, although this quantity may not fully show localized discrepancies within specific frequency regions. 

Since the results in the modest regime lie rather close to the benchmark when measured using the KL divergence, we may take the corresponding averaged value, $\mathrm{KL}_{avg} \approx 0.02$, as a practical reference point. Values larger than this threshold represent that the compared GW spectra deviate from each other significantly.

\begin{figure}[H]
  \centering
  \includegraphics[width=0.49\textwidth]{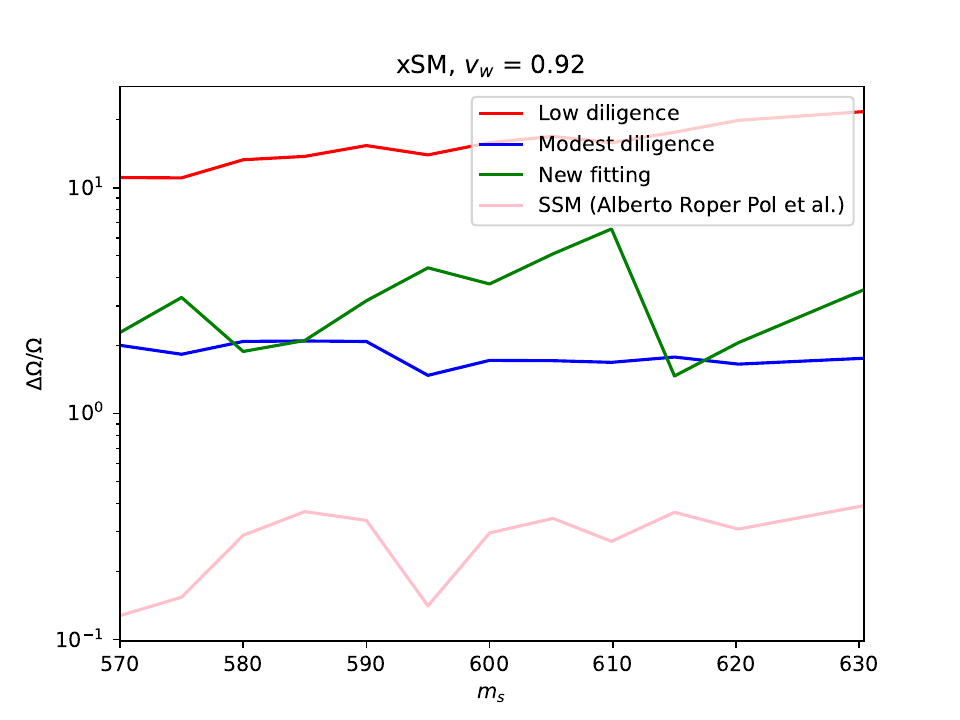}
    \includegraphics[width=0.49\textwidth]{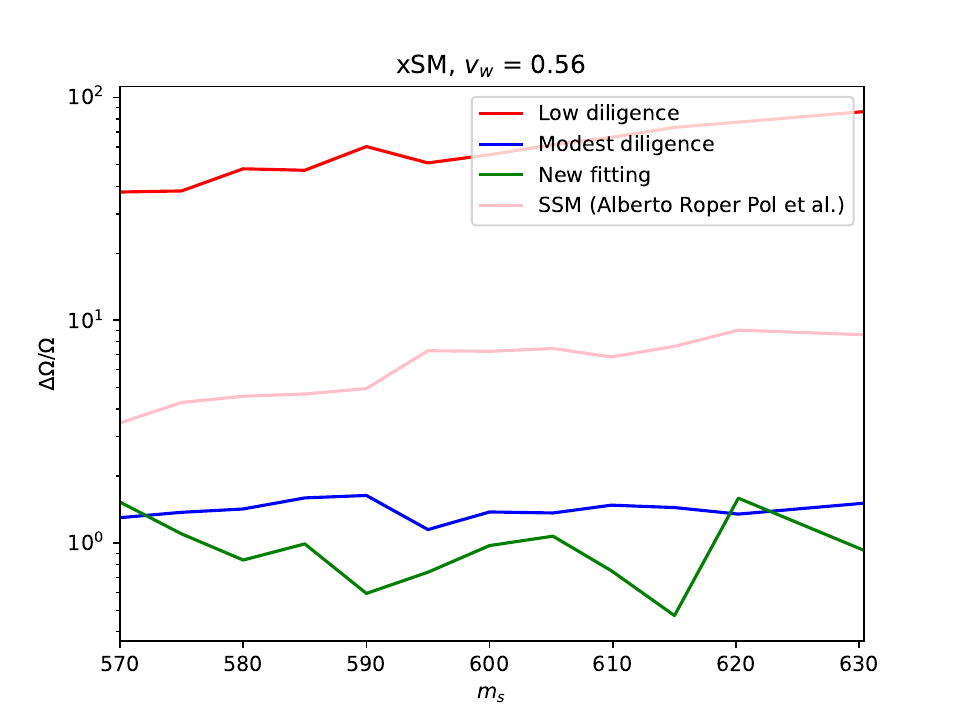}
  \includegraphics[width=0.49\textwidth]{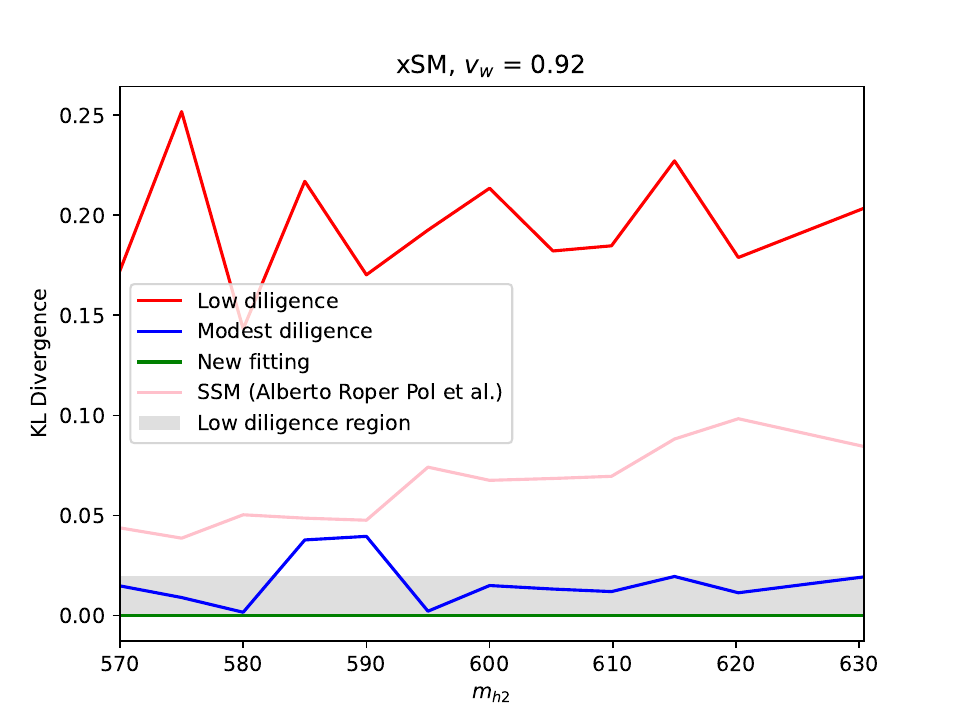}
  \includegraphics[width=0.49\textwidth]{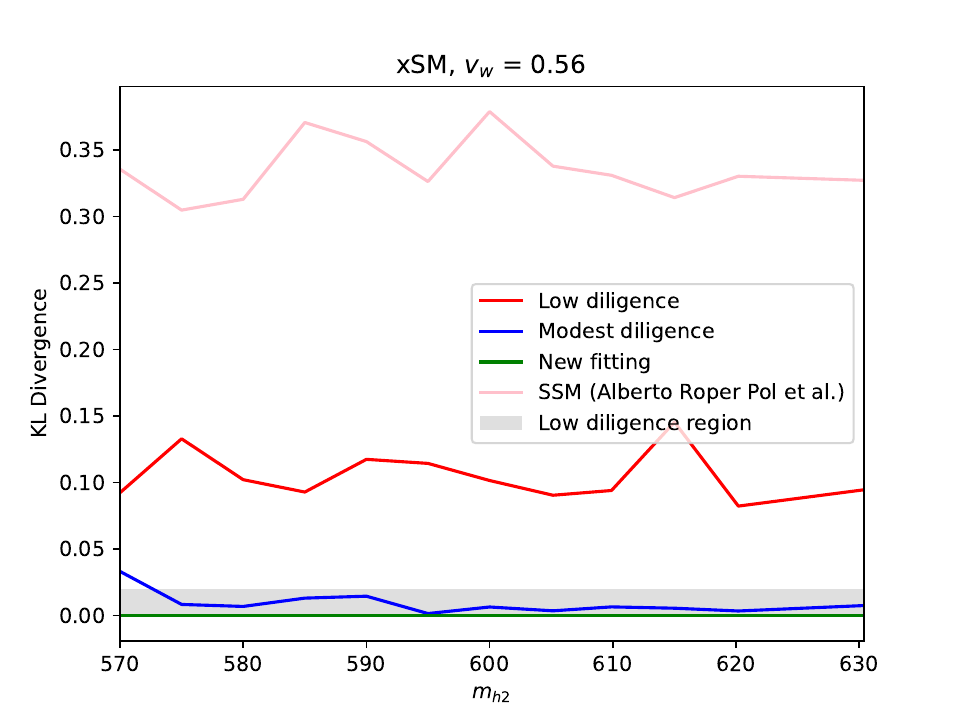}
\caption{
(Top) Comparison of the uncertainties in the peak amplitude of the gravitational wave spectra obtained using different computational approaches at various wall velocities;
(Bottom) Comparison of the KL divergence of the gravitational wave spectral shapes for different computational approaches at various wall velocities. 
}
\label{fig: GW whole compare}
\end{figure}

With the above interpretation in mind, we can now analyze how the differences in the GW spectra evolve with varying $m_{h_2}$. The corresponding results are shown in the  Fig.~\ref{fig: GW whole compare}. As can be seen, as the parameter $m_s$ varies, neither $\Delta\Omega/\Omega$ nor the KL divergence evolves smoothly, and wiggles appear in the corresponding spectra. 
The wiggles originate from numerical noise in the fitting formula inherited from the underlying tabulated input data and are not physical. 
While they can be systematically reduced by increasing the sampling density or applying controlled smoothing, doing so would require a full rescan of the parameter space using the sound shell model, which is computationally expensive and beyond the scope of the present work. Since the primary goal of our work is to compare relative trends between different calculations rather than to provide high-precision predictions, we consider the present treatment sufficient for our purposes and leave a refined refitting of the formula to future work.

From the upper panel of Fig.~\ref{fig: GW whole compare}, we observe that when the bubble wall velocity is 0.92, the peak amplitude predicted by the sound shell model closely matches that of the high-diligence result, comparing with the other three results. In addition, our new fitting formula yields results similar to those obtained under modest diligence, while the low-diligence result significantly deviates from all others.For $v_w = 0.56$, the result obtained with our new fitting formula agrees most closely with that from high diligence. The sound shell model matches the new fitting formula at small values of $m_s$, but deviates significantly from it at larger masses.

\begin{figure}[H] 
  \centering
  \includegraphics[width=0.45\textwidth]{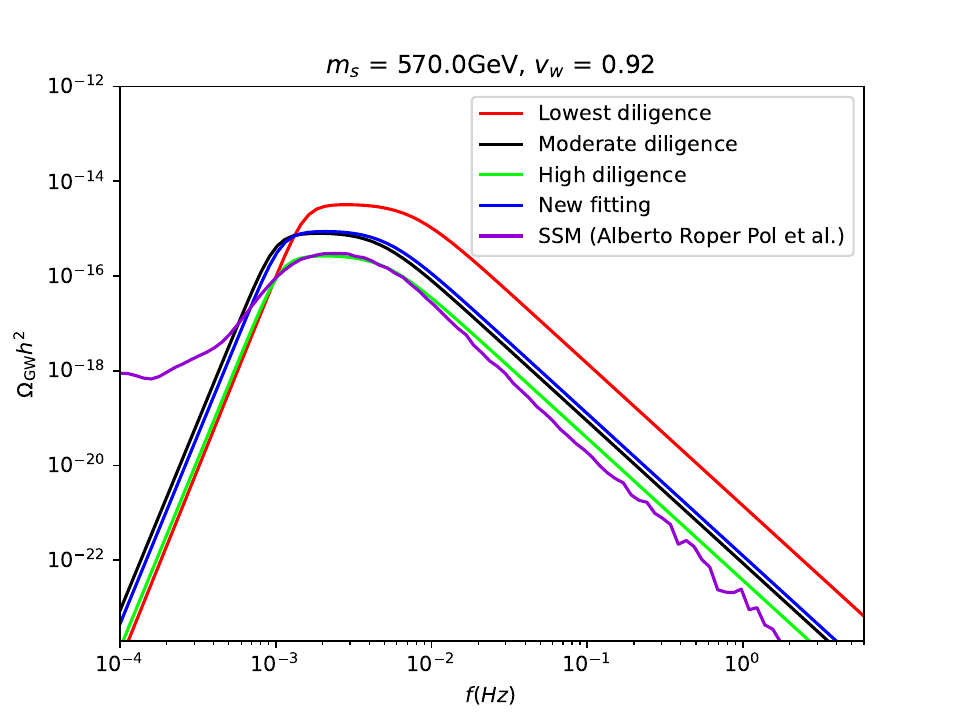}
   \includegraphics[width=0.45\textwidth]{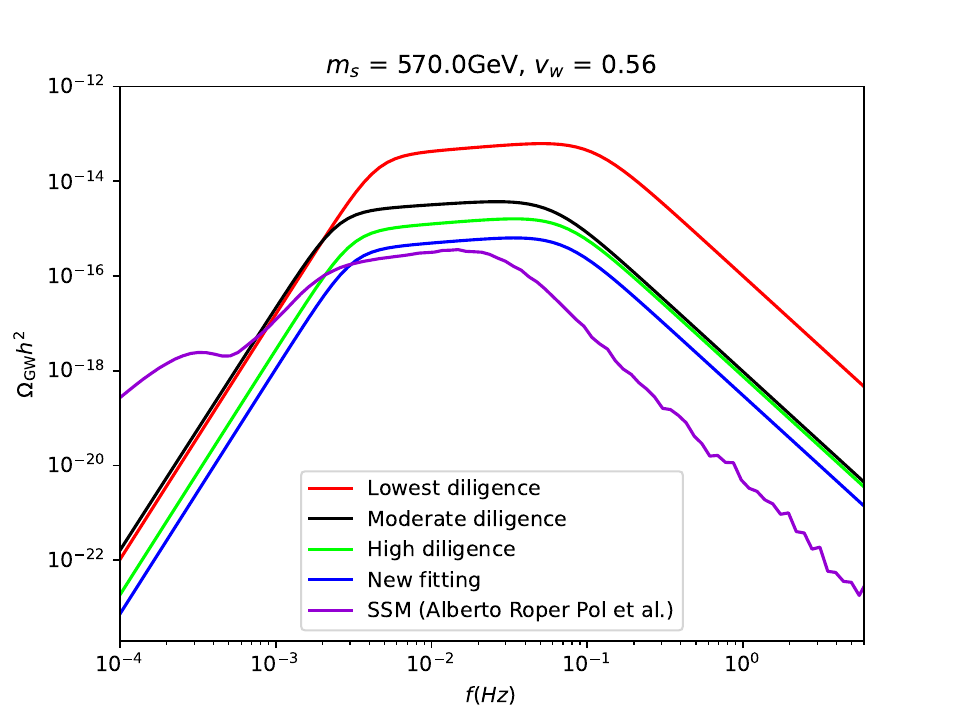}
  \includegraphics[width=0.45\textwidth]{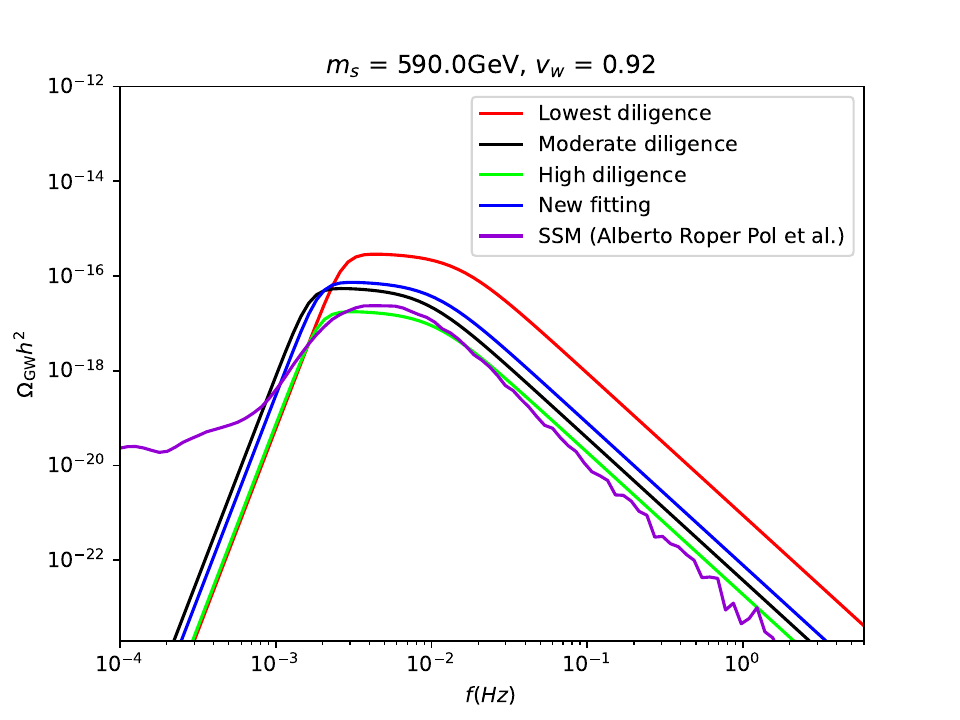}
   \includegraphics[width=0.45\textwidth]{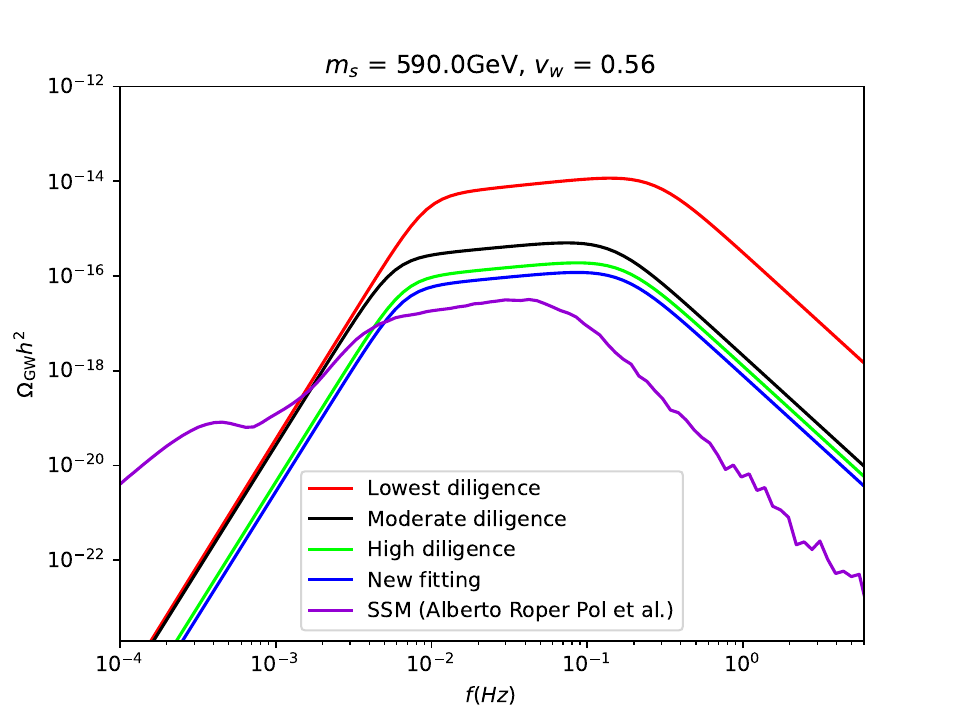}
  \includegraphics[width=0.45\textwidth]{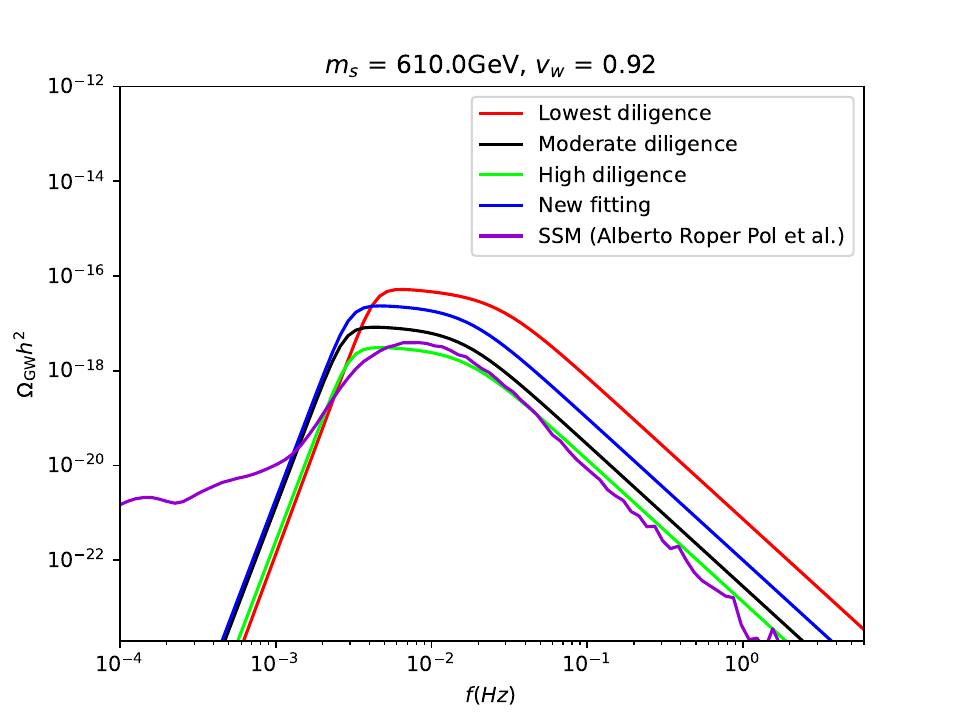}
  \includegraphics[width=0.45\textwidth]{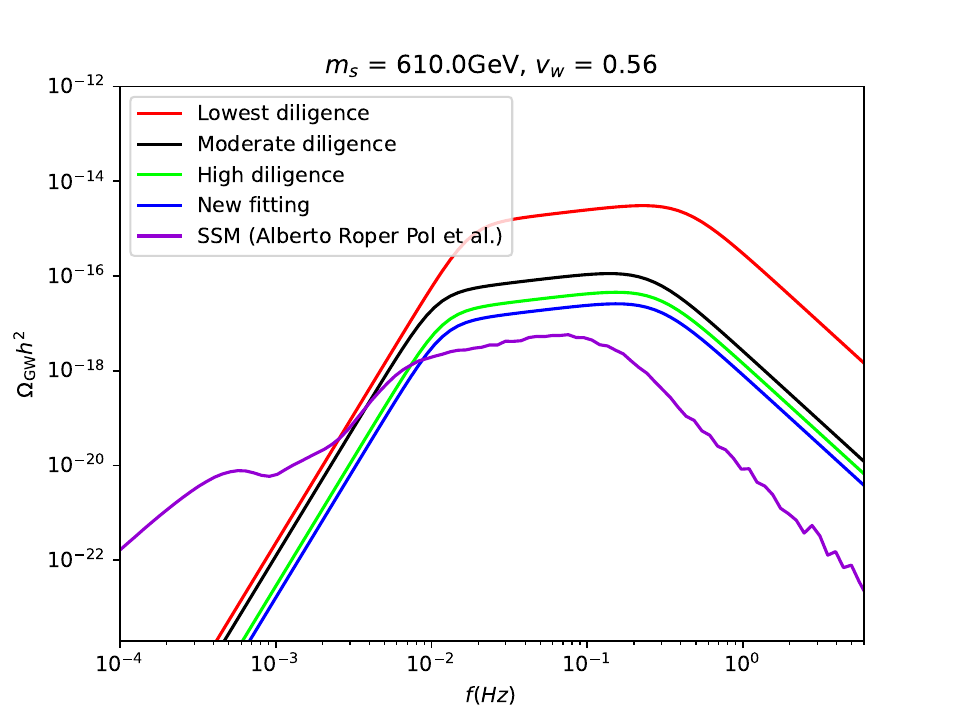}
    \includegraphics[width=0.45\textwidth]{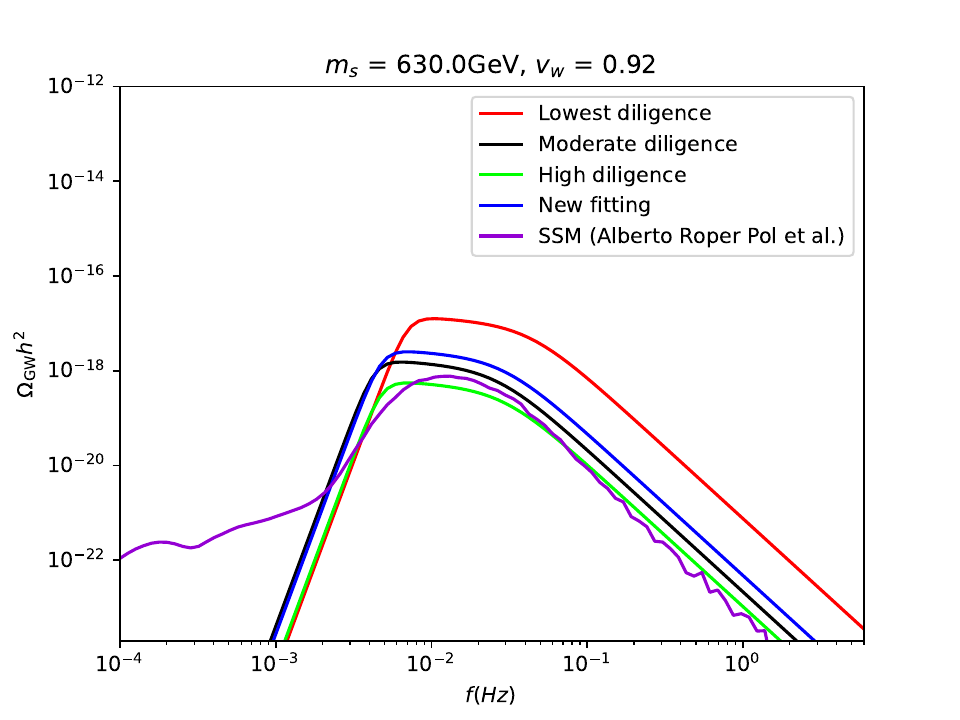}
   \includegraphics[width=0.45\textwidth]{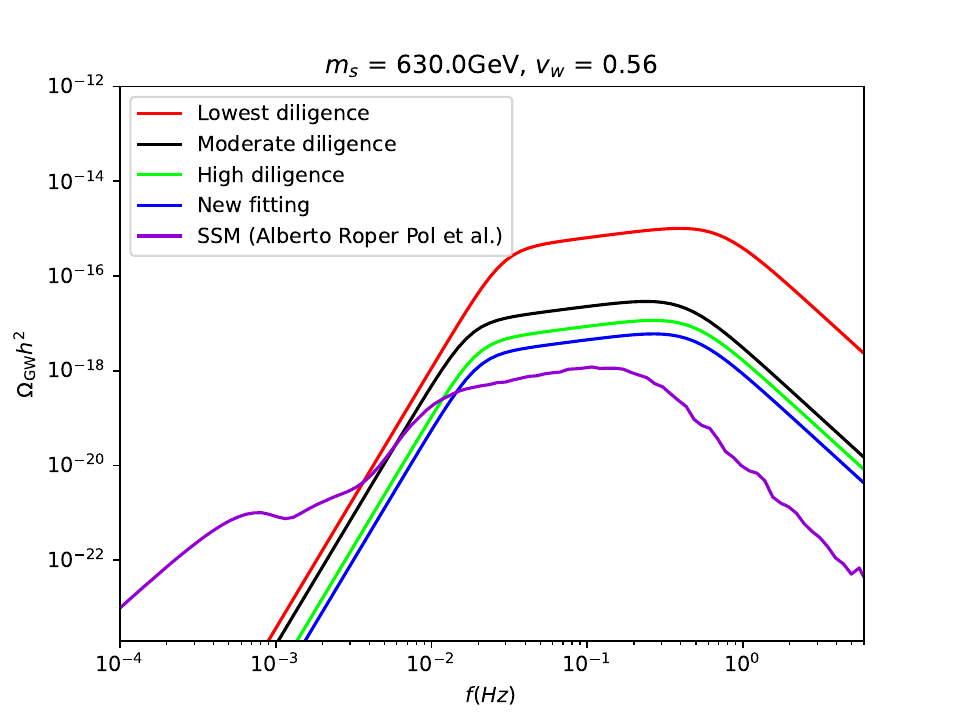}
\caption{Comparison of gravitational wave spectra at the benchmark points of the xSM from Table \ref{tab: BP}. We have considered four different scalar masses as denoted on top of each plot. We also assumed two different wall velocities $v_w = 0.92$ and $v_w = 0.56$ in the left and right panels of above figure, respectively.}
\label{fig: GW BP}
\end{figure}

From the lower panel of Fig.~\ref{fig: GW whole compare}, we find that, regardless of velocity, the spectral shapes obtained from high diligence, modest diligence, and our new fitting formula are nearly identical, except in some small mass region. In contrast, the results from the sound shell model and the lowest diligence level differ significantly from those obtained with high diligence. For the former, this is likely because the $\Delta$ function in the sound shell model is computed exactly, leading to the expected behavior $k^3$ in the low-frequency regime. This differs from the behavior predicted by lattice-based fitting formulas, which accounts for the large KL divergence observed. To investigate this discrepancy for the latter, we plot the gravitational wave spectra at several benchmark points in Fig.~\ref{fig: GW BP}.

We see in Fig.~\ref{fig: GW BP} that even though the changes in diligence result in modest changes in the thermal parameters $\alpha$ and $\kappa$ of the order of $\sim 5\%$.  The shape of the peak represents a large amount of sensitivity that broadens and changes its gradient. In addition, as noted by \cite{RoperPol:2023dzg} changes to the infrared part of the spectrum due to causality can produce a modification near the peak which we see for our high velocity benchmark. 

We can also see in Fig.~\ref{fig: GW BP}, we observe that the gravitational wave spectra obtained with the lowest diligence exhibit a similar overall shape to those obtained with the other two diligence levels, which aligns with our expectations. However, the peak frequency in the lowest diligence case deviates significantly from the others, resulting in a larger KL divergence. A closer inspection of Tab.~\ref{tab: BP} shows that the characteristic temperatures in the three diligence prescriptions all lie near $T_* \simeq 100~\mathrm{GeV}$ and differ only at the percent level, whereas $\beta/H$ changes much more significantly. For instance, $\beta/H$ drops from $\sim 4.1\times 10^3$ in the low-diligence case to $\sim 2.5\times 10^3$--$2.7\times 10^3$ in the moderate- and high-diligence cases. Since $\beta/H$ enters directly as an input to the shape function used to compute the GW spectrum, such variations primarily manifest as shifts in the peak frequency. In the low- and moderate-diligence, the functional form of $\beta/H(T)$ is identical, and the only difference is the choice of characteristic temperature. Because $\beta/H(T)$ decreases rapidly in this temperature range, even a small change in $T_*$ leads to a sharp reduction in $\beta/H$, and hence a noticeable shift of the peak. However, the effect of cosmic expansion is treated more carefully, the value of $\beta/H$ in the high-diligence case is slightly different from the moderate-diligence value. Then the corresponding peak frequencies remain very close to each other. 

\begin{table}[htbp]
\caption{The characteristic quantities for the benchmark point $m_s \approx 615$ GeV under different levels of diligence. 
}
\label{tab: BP}
\centering
\begin{tabular}{llllllllll}
\hline\hline
case & $T_{*}$ & $\alpha$ & $\beta/H$ & $\kappa$ & $t_{\rm sw}$ & $\Upsilon$ & $B$ & $KL$ &$\Delta \Omega / \Omega$ \\
\hline
\multicolumn{9}{l}{$v_w = 0.56$} \\
\hline
low   & 100.56 & 0.0131 & 4136.97 & 0.1317 & -   & -   & -   & 0.1454  &73.4006 \\
moderate & 99.37 & 0.0141 & 2484.10 & 0.1372 & 0.0173 & -   & -   & 0.0056  &1.4437 \\
high   & 99.84 & 0.0137 & 2716.55 & 0.1402 & -   & 0.0133 & 0.596 & 0  &0 \\
\hline
\multicolumn{9}{l}{$v_w = 0.92$} \\
\hline
low   & 100.56 & 0.0131 & 4136.97 & 0.0215 & -   & -   & -   & 0.2272  &17.6210 \\
moderate & 99.43 & 0.0140 & 2511.83 & 0.0230 & 0.0689 & -   & -   & 0.0195  &1.7807 \\
high   & 99.89 & 0.0136 & 2762.66 & 0.0214 & -   & 0.0628 & 0.534 & 0  &0 \\
\hline\hline
\end{tabular}
\end{table}

Moreover, we can see that the GW spectrum from the sound shell model represents complex infrared behavior with a secondary peak frequency for certain parameter choices. Such features cannot be captured by fitting formulas based on lattice simulation results. The use of a double broken power law is shown to be insufficient to address this issue~\cite{Guo:2024gmu}. 

However, these structures exist in simulation results. This motivates us to look for a form of $S_{\rm sw}(f)$ that more accurately reflects the true spectral shape. 
Additionally, we find that the peak value of GW spectrum predicted by the sound shell model is comparable to that obtained under modest diligence, but differs from the result under high diligence. This discrepancy arises mainly because high diligence calculations take into account the loss of sound wave energy due to reheating, an effect that is not included in the sound shell model.

\section{Effects of the Speed of Sound on the New Sound Shell Model}
\label{sec:sec-sp-soun}
The fluid velocity profile is important in the sound shell model and influences the produced gravitational waves spectrum from phase transition. This profile is highly sensitive to the approximations adopted for the equation of state of fluid. In the bag model, the speed of sound in both phases is assumed to be a constant $1/\sqrt{3}$~\cite{Espinosa:2010hh}, whereas in the $\mu\nu$ model, it is determined from the full effective potential~\cite{Giese:2020znk, Giese:2020rtr}.
Since the speed of sound explicitly enters the fluid equations of motion, different values naturally lead to different fluid configurations, thereby affecting the gravitational wave from sound waves. In the following, we investigate how variations in the speed of sound affect the predictions of the sound shell model.

The fluid profile influences the gravitational wave signal through its impact on the correlation function of the fluid velocity field, $ E_{\rm kin}$. The calculation of $E_{\rm kin}$ requires the extraction of normalized energy fluctuations \cite{Hindmarsh:2019phv}
\begin{equation}
  \lambda = \frac{e - \bar{e}}{\bar{w}} \, .
\end{equation}
In bag model, we can relate this to the transition strength $\alpha_b$:
\begin{align}
  \lambda &= \frac{e - \bar{e}}{\bar{w}} \notag\\
   &= \frac{e - \frac{3}{4}\bar{w} - \epsilon}{\bar{w}} \notag \\
   &= \frac{e}{\bar{w}} - \frac{3}{4} - \frac{\epsilon}{\bar{w}} \notag \\
   &= \frac{e}{\bar{e}}\frac{\bar{e}}{\bar{w}} - \frac{3}{4} - \frac{3}{4}\alpha .
\end{align}
In the second equality, we have used the equation of state of the bag model \cref{eq: bag model} and assumed that the average enthalpy and energy densities are close to their values in the symmetric phase. In the last equality, we have used the definition of transition strength \cref{eq: alpha bag}: $\alpha_b = \frac{4 \epsilon}{3 w_{s}} = \frac{4 \epsilon}{3 \bar{w}}$.

\begin{figure}[H]
  \centering
  \includegraphics[width=0.49\textwidth]{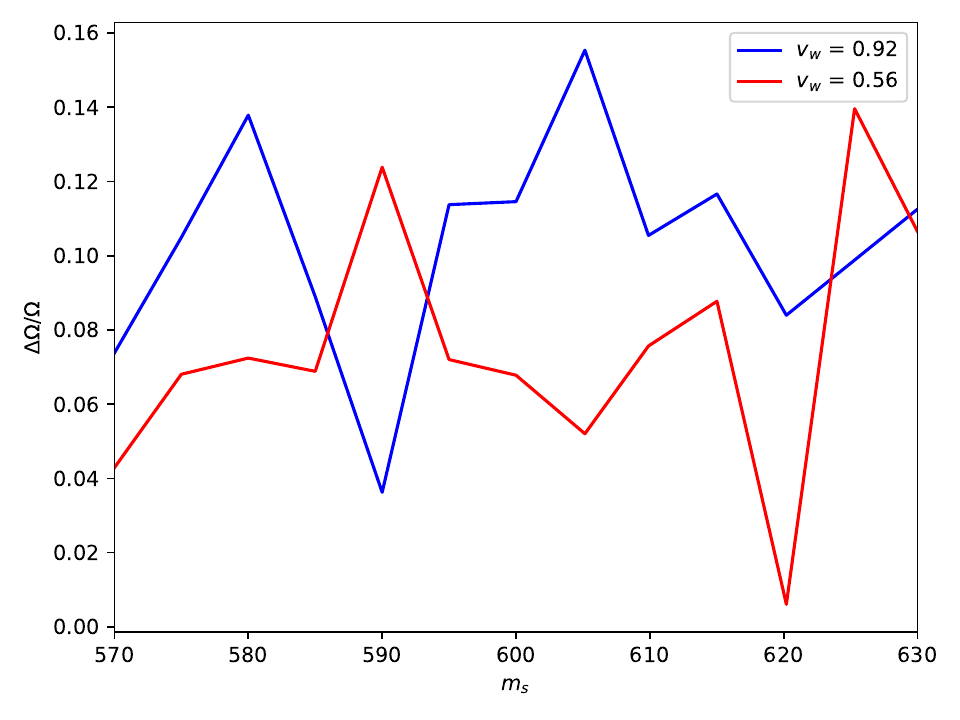}
  \includegraphics[width=0.49\textwidth]{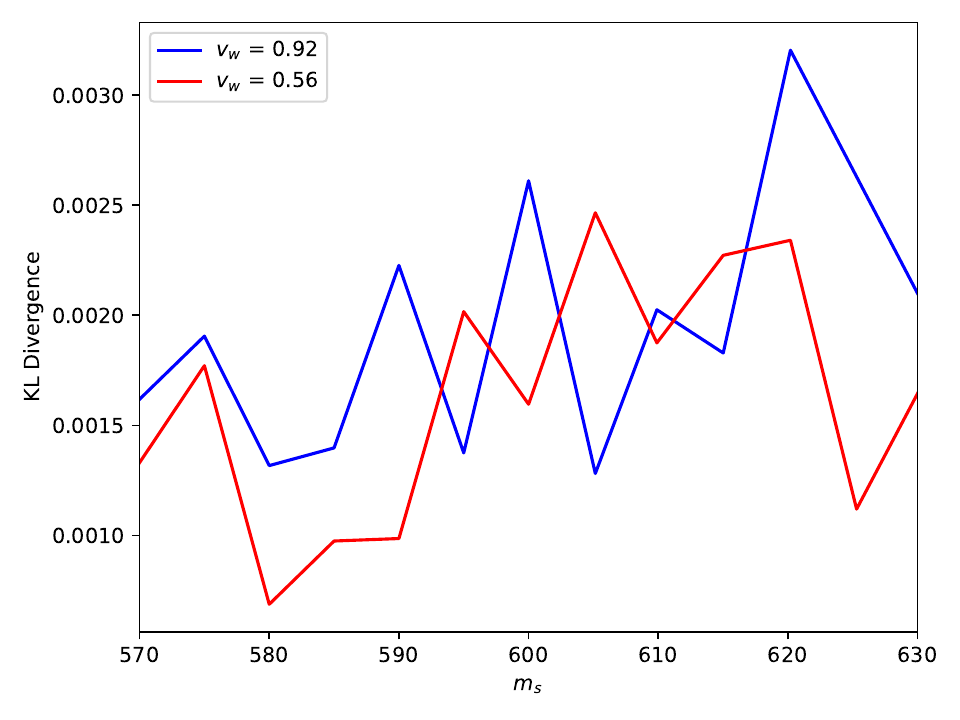}
\caption{(Left) Comparison of the uncertainties in the peak amplitude of the gravitational wave spectra resulting from the $\mu \nu$ model and the bag model at various wall velocities;
(Right) Comparison of the KL divergence between the gravitational wave spectral shapes predicted by the $\mu \nu$ model and the bag model at various wall velocities.}
\label{fig: munu_bag_gw_compare_all}
\end{figure}

For $\mu \nu$ model, the difference in the equation of state prevents us from directly applying the above results, and so a re-derivation based on the specific equation of state is required. 
To simplify the final expression, we first rewrite the equation of state in the $\mu \nu$ model \cref{eq: mu_nu model} in terms of the enthalpy density
\begin{align}
  e_s = \frac{\mu-1}{\mu}w_s + \epsilon, ~~~~p_s = \frac{w_s}{\mu} - \epsilon, \notag \\
  e_b = \frac{\nu-1}{\nu}w_b, ~~~~p_b = \frac{w_b}{\nu}.
\end{align}
Substituting the above expressions into the definition of $\lambda$, we can obtain the following \cite{Hindmarsh:2019phv}
\begin{align}
  \lambda &= \frac{e - \bar{e}}{\bar{w}} \notag\\
  &= \frac{e - \frac{\mu-1}{\mu} \bar{w} - \epsilon}{\bar{w}} \notag\\
  &= \frac{e}{\bar{e}}\frac{\bar{e}}{\bar{w}} - \frac{\mu -1}{\mu} - \frac{\epsilon}{\bar{w}}, 
\end{align}
where we again assume that the average enthalpy and energy densities are close to their values in the symmetric phase. To relate the terms involving $\epsilon$ to the phase transition strength, we can examine the definition of $\alpha_{\theta}$ using Eq.~(\ref{eq:alpha-theta})
\begin{align}
  \alpha_{\theta} &= \frac{\theta_s - \theta_b}{3w_s}, \notag \\
  &=\frac{\left(e_s - e_b\right) + \left(\nu-1\right)\left(p_b-p_s\right)}{3w_s}, \notag \\
  &=\frac{\frac{\mu-1}{\mu}w_s +\epsilon - \frac{\nu-1}{\nu}w_b + (\nu-1)\left[\frac{w_b}{\nu} - \frac{w_s}{\mu} + \epsilon\right]}{3w_s}, \notag \\
  &=\frac{1}{3}\left(\frac{\mu-1}{\mu} - \frac{\nu-1}{\mu}\right) + \nu \frac{\epsilon}{3w_s}.
\end{align}
In the second equality, we have used the fact that $\nu -1 = \frac{1}{c_{s, b}^2}$. As can be seen, different models and their respective parameters lead to different values of $\mu$ and $\nu$, indicating that the $\mu\nu$ model captures more details of the underlying physics. 
Moreover, it is evident that the bag model is a special case of the $\mu \nu$ model, corresponding to $\mu = 4$ and $\nu = 4$. 
Once the fluid profiles are obtained, they can be incorporated into the sound shell model to compute the resulting gravitational wave spectra. The Fig.~\ref {fig: GW BP} scan results are shown in Fig.~\ref{fig: munu_bag_gw_compare_all}, where the small wiggles are mainly attributed to numerical integration errors. Note that the relative error is large, but this is an artifact of the KL divergence being very small. 

The way in which the peak amplitude is affected by the equation of state is represented in the left panel of Fig.~\ref{fig: munu_bag_gw_compare_all}, where we show the relative difference $\Delta\Omega/\Omega$ between the model $\mu\nu$ and the bag model. 
Overall, this difference is significantly smaller than the case discussed in the last section, indicating that the uncertainty associated with the choice of equation of state is subdominant compared to the choice of diligence. For the larger wall velocity, $v_w = 0.92$, the quantity $\Delta\Omega/\Omega$ exhibits a mild upward trend as $m_s$ increases, whereas for the smaller wall velocity, $v_w = 0.56$, it shows an oscillatory behaviour as a function of $m_s$ but remains below the $v_w = 0.92$ curve for most of the parameter range. 
This pattern suggests that the mismatch induced by the different sound speeds in the two equations of state becomes less important at smaller wall velocities.

\begin{figure}[H] 
  \centering
  \includegraphics[width=0.49\textwidth]{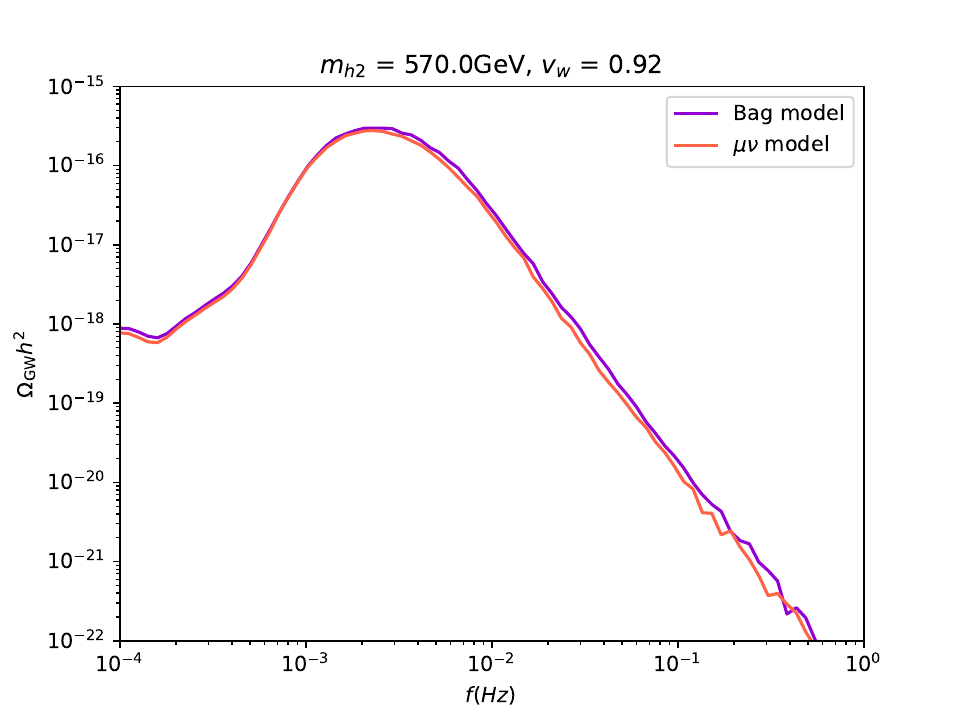}
   \includegraphics[width=0.49\textwidth]{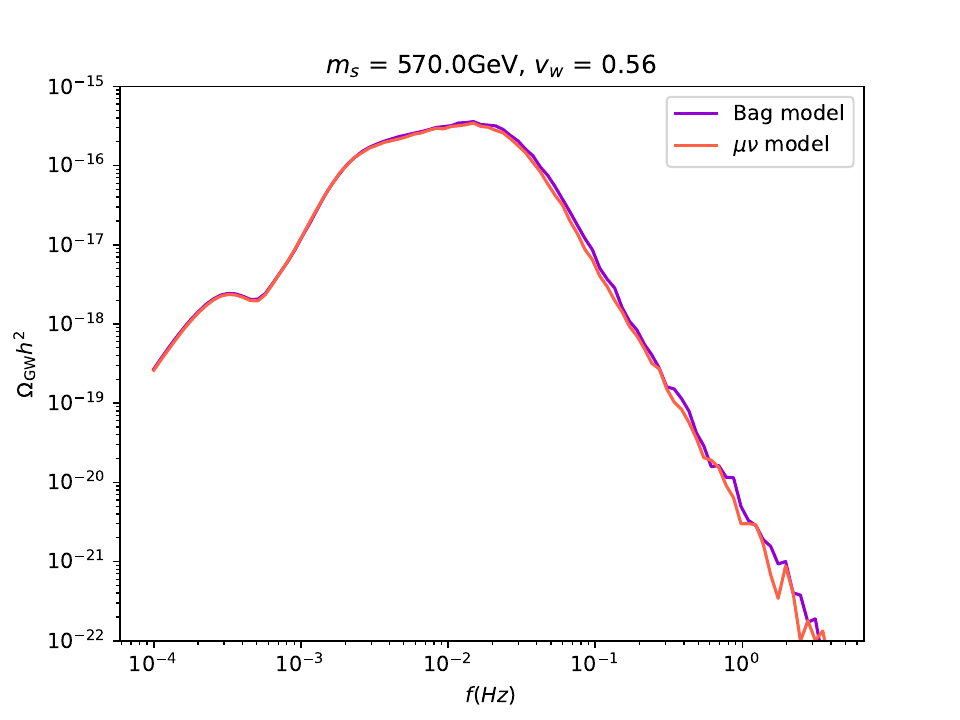}
  \includegraphics[width=0.49\textwidth]{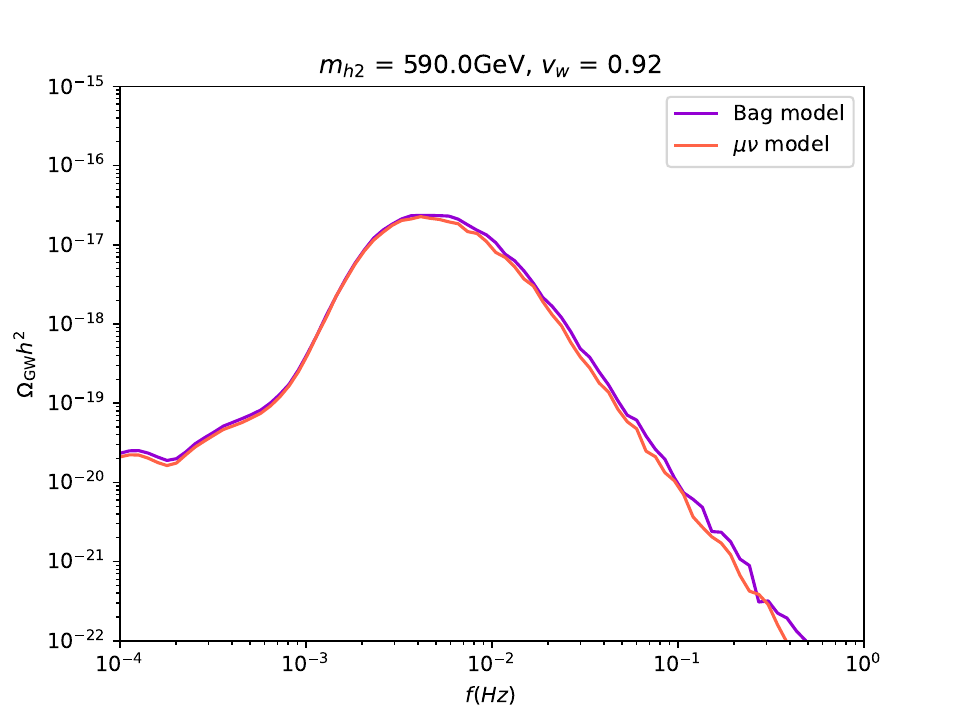}
   \includegraphics[width=0.49\textwidth]{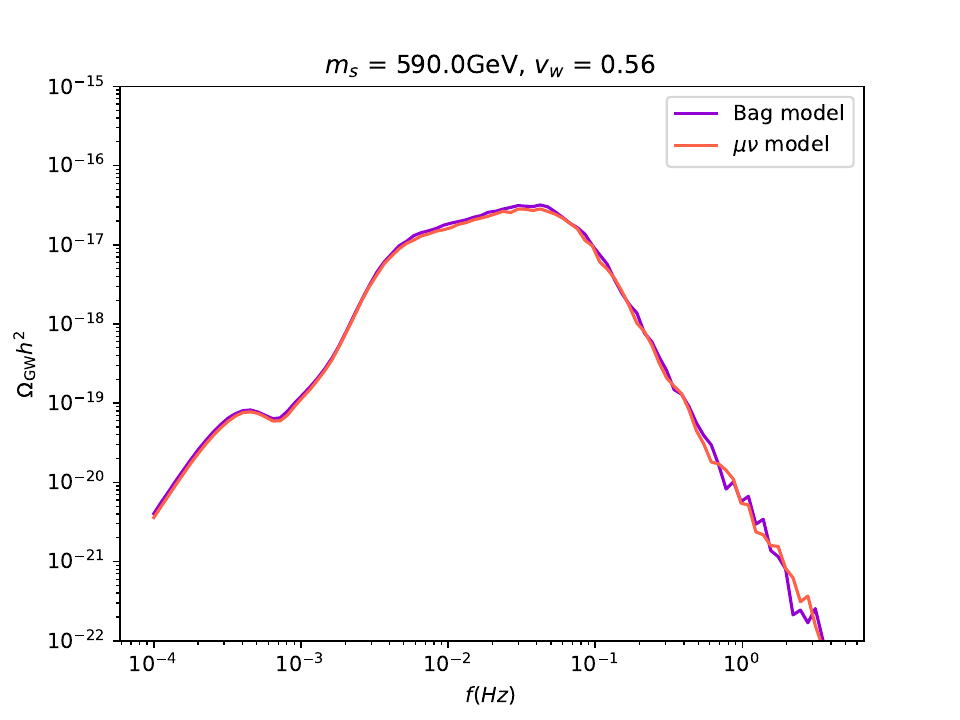}
  \includegraphics[width=0.49\textwidth]{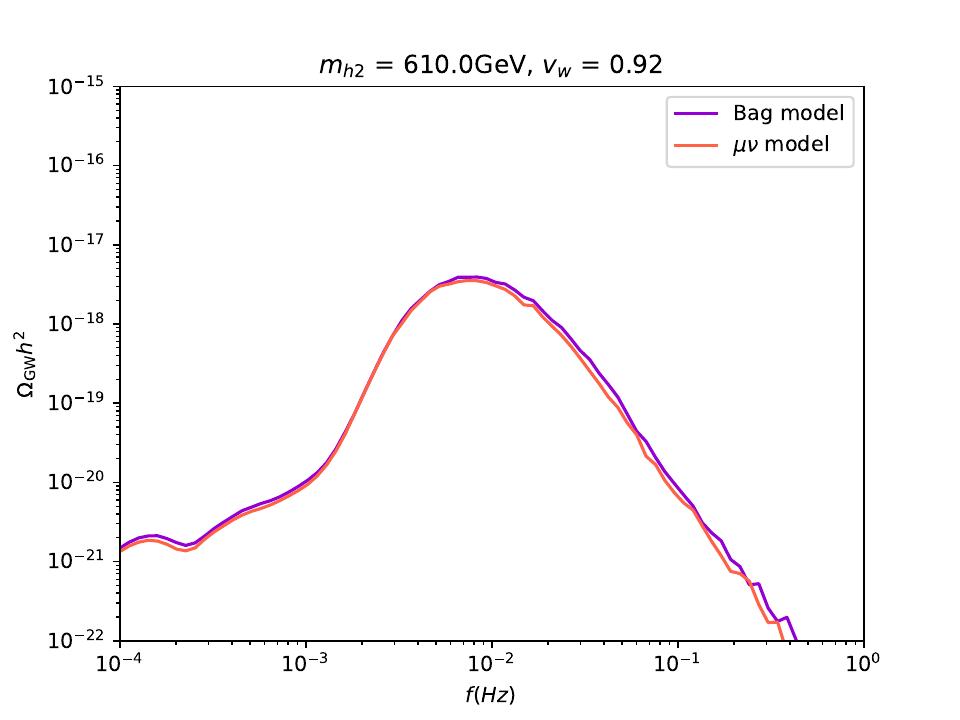}
   \includegraphics[width=0.49\textwidth]{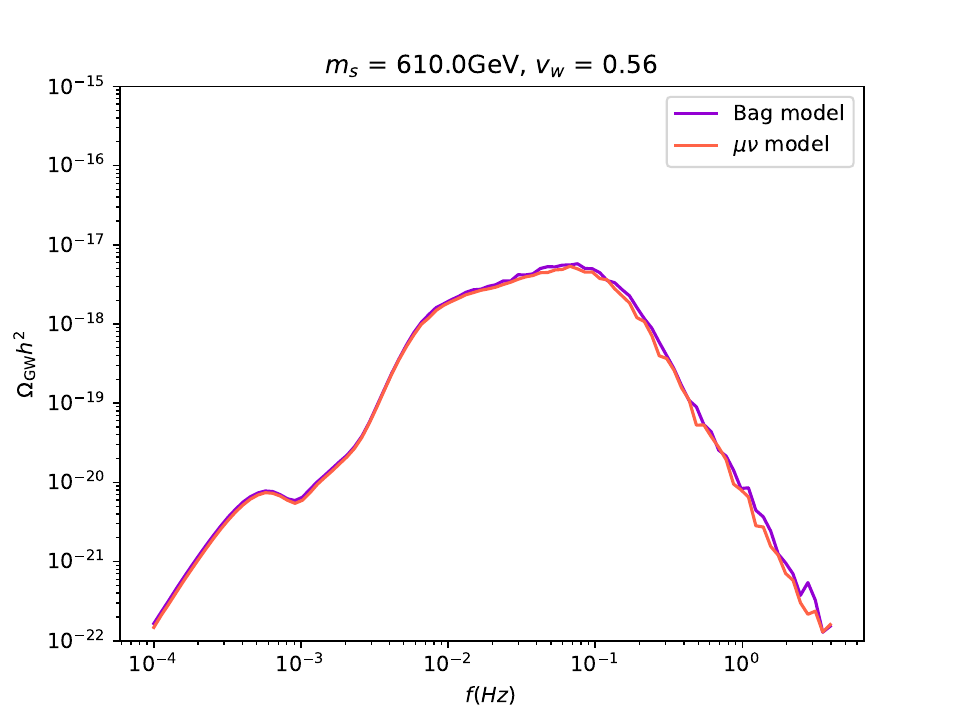}
  \includegraphics[width=0.49\textwidth]{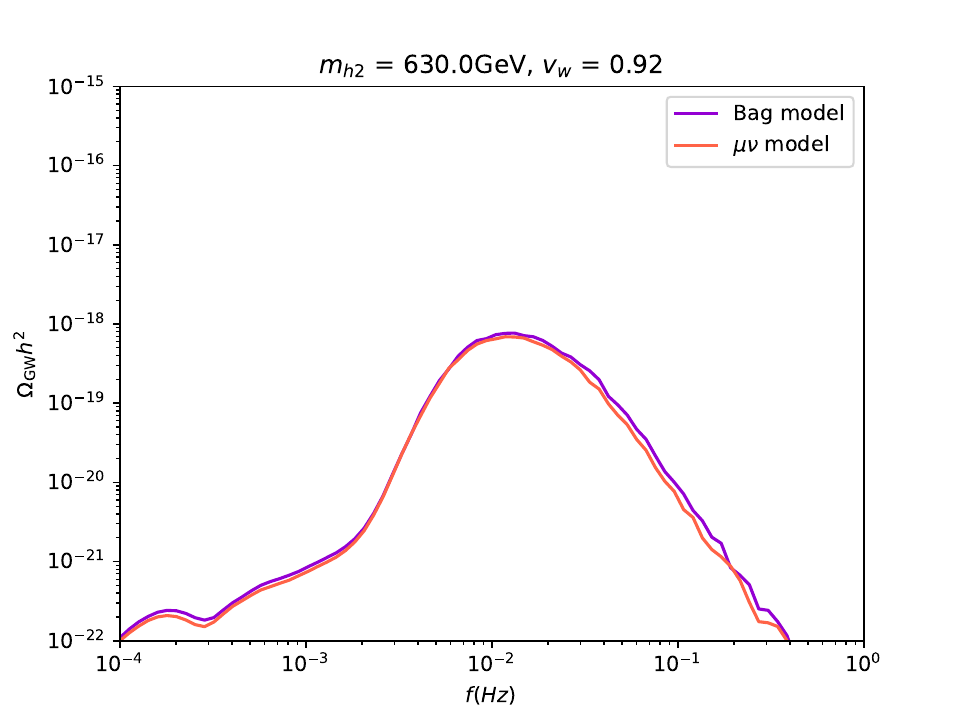}
  \includegraphics[width=0.49\textwidth]{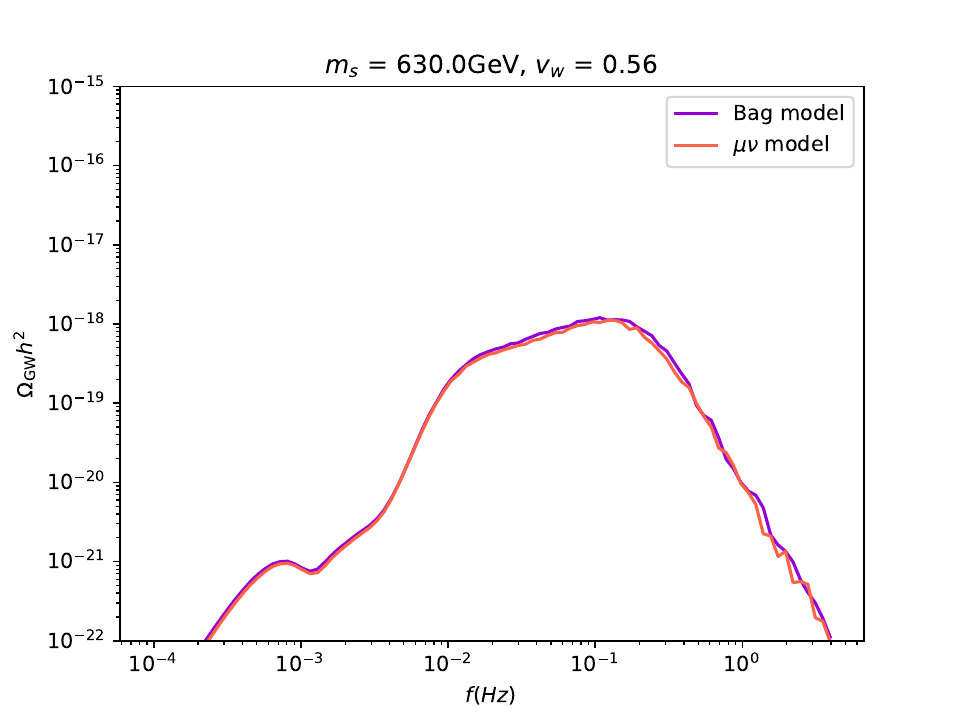}
\caption{The comparison plot of gravitational wave spectra resulting from the $\mu \nu$ model and the bag model at the benchmark points of the xSM.}
\label{fig: munv_bag_GW BP}
\end{figure}

The right panel of Fig.~\ref{fig: munu_bag_gw_compare_all} presents the KL divergence between the spectra obtained from the $\mu\nu$ model and the bag model. 
For all benchmark points considered, the KL values remain well below $0.02$, which implies that the shapes of the resulting gravitational wave spectra are practically indistinguishable. We present several benchmark points in Fig.~\ref{fig: munv_bag_GW BP} to provide a more clear visual illustration of these effects.

\begin{figure}[htbp!]
  \centering
  \includegraphics[width=0.49\textwidth]{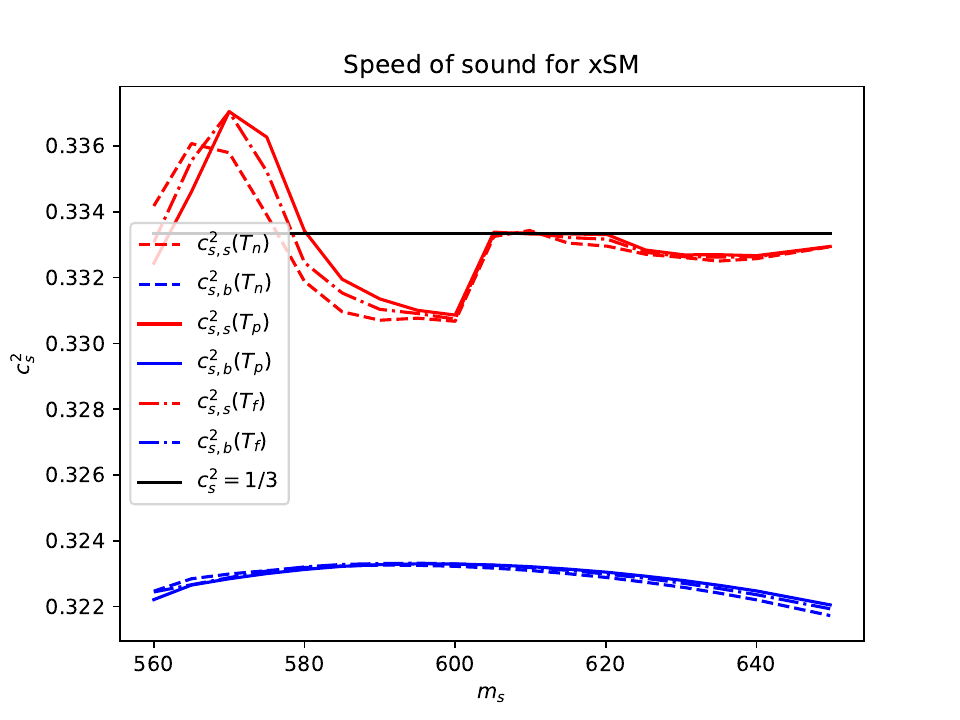}
\caption{The variation of the squared speed of sound in the symmetric and broken phases with $m_{s}$ at different temperatures.}
\label{fig: sound speed}
\end{figure}

The small discrepancies observed above primarily originate from the fact that the sound speeds in the symmetric and broken phases differ only mildly. 
Consequently, the fluid profiles computed from different equations of state remain very similar, which in turn leads to only minor differences in the resulting gravitational wave spectra. This is shown in Fig.~\ref{fig: sound speed} the sound speeds in the two phases as functions of $m_s$. Within the parameter range considered, the sound speed in the symmetric phase varies only between $c_{s,s}^{2}\in[0.330,\,0.336]$, while in the broken phase it remains close to $c_{s,b}^{2}\simeq 0.323$, exhibiting a slight downward trend as $m_s$ increases. 
This indicates that achieving a significantly larger $c_s$ deviation would require simultaneously large values $\mu_s^2$, and $\lambda_{hs}$, which would in turn invalidate the assumption that one loop corrections to the effective potential are sufficient. In this sense, sizeable deviations in the sound speed are difficult to realize within the simplified parameter space adopted here. However, if the model includes particles with masses comparable to the phase-transition temperature contributing to relativistic degrees of freedom, the sound speed can deviate from the initial expectation~\cite{Giese:2020znk,Si:2025vdt}.

\begin{figure}[H] 
  \centering
  \includegraphics[width=0.49\textwidth]{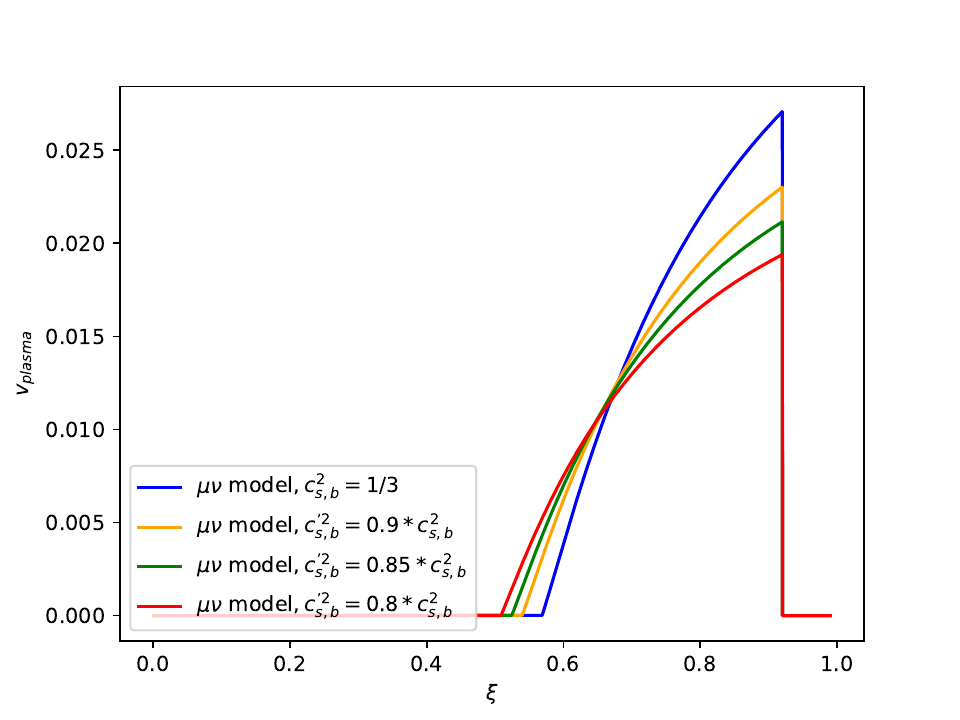}
   \includegraphics[width=0.49\textwidth]{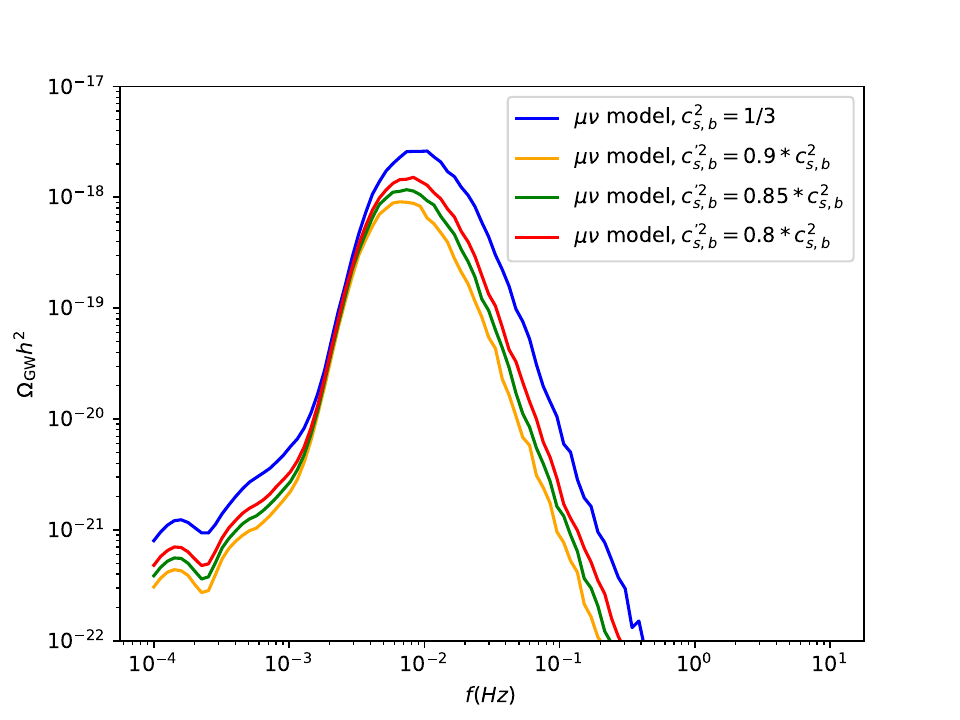}
  \includegraphics[width=0.49\textwidth]{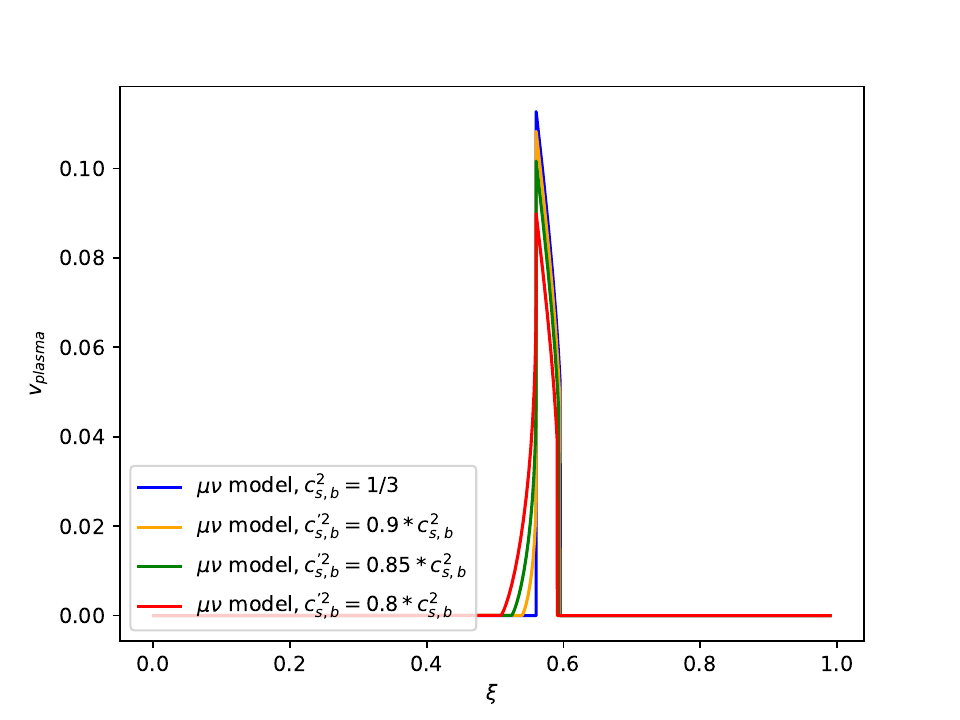}
   \includegraphics[width=0.49\textwidth]{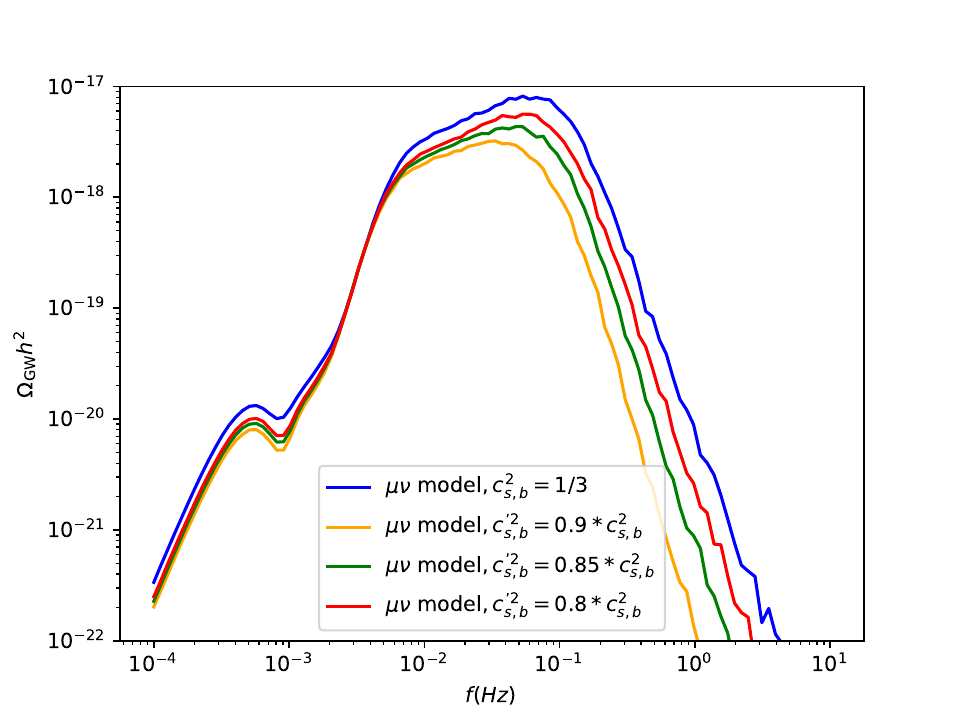}
\caption{The dependence of the fluid profiles and the sound shell model gravitational wave signal on the speed of sound in the broken phase. We considered $v_w =0.56$ and $v_w =0.92$ in the upper and lower panels, respectively.}
\label{fig: munv_bag_cs}
\end{figure}

Based on previous work, it is possible that the speed of sound in a concrete model can change up to order $\Delta c_s^2\sim 6\%$ \cite{Tenkanen:2022tly} even in simple extensions of the Standard Model. This motivates considering the effect of a larger change in the speed of sound. Therefore, to quantify the uncertainty associated with the speed of sound variation in the improved sound shell model, we select a representative benchmark point ($m_s \approx 605~\mathrm{GeV}$, $\lambda_{hs} = 5$) and manually vary the speed of sound in the broken phase to assess its impact. The results are shown in the Fig.~\ref{fig: munv_bag_cs}.

We observe that for $v_w = 0.92$, decreasing the sound speed in the broken phase suppresses the fluid velocity profile and changes the left zero of the profile toward a smaller location. However,  the situation is more complex for $v_w = 0.56$. When $c_{s.b}^2 = 1/3$, the wall velocity is slightly below the sound speed ($c_{s,b} \simeq 0.577$), and the fluid shows a deflagration type profile. 
While we reduce the speed of sound, this condition is not valid any more, and the fluid profile transitions into a hybrid profile. 
A similar transition also occurs when $v_w$ is close to $v_J$. 
This indicates that the fluid profiles near such critical points are highly sensitive to the speed of sound in the broken phase. Consequently, estimating the effect of the speed of sound on the sound shell model depends on the choice of wall velocity. This shows the significant uncertainty introduced by the bubble wall velocity in predicting the gravitational wave spectrum.

 \noindent From the produced gravitational wave spectra, we find that if the speed of sound of broken phase reduces, then the entire spectrum moves  downward. In particular, at smaller wall velocities, the ultraviolet part of the spectrum is suppressed more strongly, which is likely related to the change in the fluid propagation mode.

\section{Summary and Conclusions}
\label{sec:summ-conc}

In this paper, we have considered three main points on gravitational production from first order phase transitions.
First, we have probed how different levels of computational precision in the bubble nucleation rate and nucleation and percolation temperature and ``diligence" in the calculation of thermal parameters can affect the whole spectrum from FOPT for a given model. We consider three cases of diligence over different frequencies that their peak frequencies have been studied in Ref. \cite{Guo:2021qcq}. 
A minimal diligence case uses the nucleation temperature $T_n$, a saddle-point nucleation rate, the bag equation of state, and $\kappa_{\rm sw}$ taken from bag-model hydrodynamic model. 
A moderate diligence case can substitute $T_n$ with the percolation temperature $T_p$. It includes the finite sound-wave lifetime via $R_*$ and the root mean square fluid velocity $U_f$. 
The highest level of diligence computes $h$, $T_f$, $R_*$, the bubble number density, and $\beta/H$ from the full nucleation evolution in an expanding universe. Then it leads to $\kappa_{\rm sw}$ and $U_f$ from hydrodynamic that considers finite-lifetime and reheating suppression analytically. These will be used as inputs and will be fed into the complete velocity profile to compute the sound shell model that can be used for the final calculation of GW spectrum, using the $\mu \nu$ model for the equation of state.

Moreover, we have studied how different modeling of sound shell model can change the predicted spectra using two features: peak-height shifts and a KL-based shape distance. 
Going from low to moderate diligence can change the peak by about an order via temperature and the duration of acoustic effects in the xSM model. 
The high diligence case keeps the peak nearly unchanged. However, it modifies the shape of spectrum when details profiles and non-bag hydrodynamics are included. 
We also fix the infrared behavior of the sound-shell model by replacing a delta-function (not using $k^9$) with an integral form that gives causal scaling of $k^3$. 
The sound-shell prediction can match with the high-diligence result especially when  the finite lifetime of source is considered precisely. However, the former $k^9$ infrared scaling of the original shell-model formulation is corrected to the causally required $k^3$ once the unequal-time correlator is calculated based on the results in Refs.~\cite{Sharma:2023mao,RoperPol:2023dzg}.

We compare different approaches with the highest diligence case for a singlet extension of the Standard Model. 
The spectrum peak frequency and amplitude can shift if one uses the nucleation temperature $T_n$. 
Depending on the choice of $T_p$ or $T_f$ there will be some discrepancy between the predicted GW spectrum. 
This shows the importance of choosing the temperature that treats the FOPT in the best way that describes the FOPT. 
In the moderate-diligence case the peak amplitudes are within $\sim10\%$ of the highest diligence case.  
The amplitude and the peak frequency in the lowest diligence case have a large deviation in comparison to other spectrum.
The $\mu\nu$ EOS enhances the kinetic energy fraction. It also shifts the spectral shape relative to the bag model. This fact represents a temperature dependence in the speed of sound and enthalpy.

Using the $\mu\nu$ model as a modification of the bag equation of state we have shown small deviations from the broken-phase sound speed from $c_{s,b}^2=1/3$ can be affected by bubble-wall matching. 
This causes an order one modification in boundary velocities and enthalpies when one assumes $|q|\sim10^{-2}$. This shows itself in velocity/enthalpy profiles in all deflagration, hybrid, and detonation regimes. 
These deviations are small in an xSM benchmark model. 
Varying $c_{s,b}^2$ when other parameters are fixed modifies the GW spectrum. Smaller $c_{s,b}^2$ values can reduce the amplitude and modify the double peak feature of GW spectrum. 
As a consequence, the spectrum depends on an additional degree of freedom beyond $(\alpha,\beta/H,v_w,T_*)$.
 This implies precise GW spectrum requires realistic equations of state and full velocity profiles. Also, it can help to constrain $c_s$ and the response in $\mathcal{F}(\phi,T)$ as model parameters.

Finally, we assumed the xSM model as a benchmark and investigated different impacts of dilligence over different frequencies of cosmological GW from FOPT. 
Also, there is a distinguishable signal between considering the bag model and its extensions like $\mu \nu$ model. In addition, we compared previous fit with the low frequency tail of GW spectrum from FOPT. Depending on the scale of phase transition any of diligence scenario can be important for nanoHz PTA experiments and miliHz regime for the space based detectors.
In future GW experiments our analysis will be useful for concrete theoretical predictions that improves the parameter space probe of FOPT. Moreover, it helps us to distinguish among various BSM scenarios based on the details of shape and amplitude of the primordial gravitational waves that we measure at current and future detectors. If a random gravitational wave signal is detected from a phase transition in the early universe, then one can identify the detailed fluid dynamics and equation-of-state based on the modeling in the frequency pattern of waves.

\acknowledgments 
We would like to thank Kuver Sinha for insightful discussions at early stages of this manuscript. 
F.H. thanks Rouzbeh Allahverdi, Nicolas Bernal and Amitayus Bhanik for useful discussions. 
He is supported by Homer Dodge postdoctoral fellowship. 
He is thankful to the organizers of workshop of Center for Theoretical Underground Physics and Related Areas (CETUP* - 2025), The Institute for Underground Science at Sanford
Underground Research Facility (SURF), Lead, South Dakota 
for their hospitality and financial support. He also thanks the organizers of the Mitchell
Conference in May 2025 at Texas A \& M University for their hospitality and support during this project. GW acknowledge the STFC Consolidated Grant ST/X000583/1. Some paragraphs of this manuscript were written in Chinese then translated using AI. We then modified the translation into our own words, but we acknowledge the use of AI in preparing this manuscript.


\bibliography{biblio.bib}
\end{document}